\documentclass[iop,apj]{emulateapj}
\usepackage{apjfonts,graphicx}
\usepackage{hyperref,url}

\newcommand{\Msun}{\,M_{\sun}}
\newcommand{\Lsun}{\,L_{\sun}}
\newcommand{\Mpc}{\,\mathrm{Mpc}}
\newcommand{\kpc}{\,\mathrm{kpc}}
\newcommand{\Gyr}{\,\mathrm{Gyr}}
\newcommand{\SFR}{\mathrm{SFR}}
\newcommand{\sSFR}{\mathrm{sSFR}}
\newcommand{\feh}{\mathrm{[Fe/H]}}

\newcommand{\Mh}{M_{\mathrm{h}}}
\newcommand{\Mg}{M_{\mathrm{g}}}
\newcommand{\MGC}{M_{\mathrm{GC}}}
\newcommand{\fb}{f_{\mathrm{b}}}
\newcommand{\fg}{f_{\mathrm{g}}}
\newcommand{\fin}{f_{\mathrm{in}}}
\newcommand{\tdep}{t_{\mathrm{dep}}}
\newcommand{\sigmamet}{\sigma_{\mathrm{met}}}

\begin{document}

\slugcomment{Accepted by ApJ}
\shortauthors{Li and Gnedin}
\shorttitle{Formation of Globular Cluster Systems}

\title{MODELING THE FORMATION OF GLOBULAR CLUSTER SYSTEMS IN THE VIRGO CLUSTER}
        
\author{Hui Li and Oleg Y. Gnedin}

\affil{University of Michigan, Department of Astronomy, Ann Arbor, MI 48109, USA;\\
    \mbox{\tt hliastro@umich.edu, ognedin@umich.edu}}

\date{\today}

\begin{abstract}
The mass distribution and chemical composition of globular cluster (GC) systems preserve fossil record of the early stages of galaxy formation. The observed distribution of GC colors within massive early-type galaxies in the ACS Virgo Cluster Survey (ACSVCS) reveals a multi-modal shape, which likely corresponds to a multi-modal metallicity distribution. We present a simple model for the formation and disruption of GCs that aims to match the ACSVCS data. This model tests the hypothesis that GCs are formed during major mergers of gas-rich galaxies and inherit the metallicity of their hosts. To trace merger events, we use halo merger trees extracted from a large cosmological N-body simulation. We select 20 halos in the mass range of $2\times 10^{12}$ to $7\times 10^{13}\Msun$ and match them to 19 Virgo galaxies with K-band luminosity between $3\times 10^{10}$ and $3\times 10^{11}\Lsun$.  To set the $\feh$ abundances, we use an empirical galaxy mass-metallicity relation. We find that a minimal merger ratio of 1:3 best matches the observed cluster metallicity distribution. A characteristic bimodal shape appears because metal-rich GCs are produced by late mergers between massive halos, while metal-poor GCs are produced by collective merger activities of less massive hosts at early times. The model outcome is robust to alternative prescriptions for cluster formation rate throughout cosmic time, but a gradual evolution of the mass-metallicity relation with redshift appears to be necessary to match the observed cluster metallicities. We also affirm the age-metallicity relation, predicted by an earlier model, in which metal-rich clusters are systematically several billion years younger than their metal-poor counterparts.
\end{abstract}


\section{Introduction}

Globular cluster (GC) systems have been found in various types of galaxies. Because of their old age and compact structure, GCs are believed to carry information on galaxy assembly history at early times \citep[e.g.,][]{brodie_strader06}. In particular, the colors and metallicities of GC systems provide a unique record of the early star formation and chemical enrichment in their host galaxies. One of the remaining puzzles is the origin of the commonly seen bimodal distribution of GC colors, within galaxies ranging from spirals to giant ellipticals.  The bimodality in color is indicative of bimodality in metallicity, which has been used to separate GCs into two subpopulations: metal-poor and metal-rich \citep{harris01, peng_etal06, harris_etal06}.

In general, GCs have systematically lower metallicity than the field stars of their host galaxy. Therefore, they must have formed earlier than the bulk of stars, at least from the chemical evolution point of view.  Early galaxies were smaller and less metal-enriched than those of today.  Motivated by this fact, we test a hypothesis that major mergers of gas-rich galaxies (which happened more frequently at high redshift, in the hierarchical galaxy formation framework) are predominantly responsible for the formation of GCs. \cite{ashman_zepf92} predicted the metallicity bimodality resulting from galaxy mergers even before observations revealed it. In their model, the two subpopulations can be produced by distinct star-forming events, which could naturally occur in hierarchical structure formation. Based on this framework, \citet[hereafter MG10]{muratov_gnedin10} modeled the metallicity distribution of Galactic GCs using the mass assembly history from a cosmological N-body simulation, coupled with observational scaling relations for galaxy stellar mass and metallicity.  This model incorporates both the formation and disruption of GCs in the progenitor galaxies of a host halo with the mass similar to the Milky Way.  The model has successfully reproduced both the bimodal metallicity distribution and the log-normal distribution of cluster mass.

During the past decade, observations of GC systems outside the Local Group have advanced significantly with the Hubble Space Telescope (HST).  A comprehensive study of galaxies in the Virgo cluster, the ACS Virgo Cluster Survey (ACSVCS), examined the photometric properties of 100 early-type galaxies, along with their GC systems.  Multi-modal GC distributions are present in all target galaxies with absolute magnitude of $-22 < M_B < -15$.  The peak metallicities of the two main modes follow a systematic (but weak) trend with galaxy luminosity, implying a possible common origin of these subpopulations. This data set gives us a good opportunity to investigate the formation of GC systems in massive elliptical galaxies. 

The GC system in the Milky Way shows only weak bimodality: 30\% of the clusters are in the metal-rich group.  In contrast, giant Virgo ellipticals have comparable numbers of red and blue clusters, and therefore, they present better tests for the origin of the metallicity distribution. 

In this paper, we extend the model of MG10 to more massive early-type galaxies and adopt the mass assembly history from a large cosmological Millennium-II (MM-II) simulation. The model is based on the calculation of the galaxy cold gas mass, the mass-metallicity relation (MMR), the cluster fraction, and the initial mass function (IMF), which we describe in Section~\ref{sec:model}. The final model is even simpler than MG10 and has only four adjustable parameters. In Section~\ref{sec:dyn}, we add the dynamical disruption of individual clusters, by considering two-body relaxation and stellar evolution.  We apply this updated model to 20 halos selected from MM-II, with total mass in the range of $10^{12} - 10^{14}\Msun$, appropriate for massive elliptical galaxies. We also investigate variants of our model in Section~\ref{sec:other_models}. In Section~\ref{sec:result} we compare the model cluster populations with the ACSVCS observations of 19 corresponding galaxies. We summarize and discuss the main results in Sections~\ref{sec:discussion} and \ref{sec:summary}.

\section{Model for Globular Cluster Formation}
  \label{sec:model}

We update the framework of the MG10 model using several independent realizations of halo assembly history and recent observational relations for galaxy stellar and gas masses.  Halo merger trees are obtained from the Millennium Database\footnote{\url{http://gavo.mpa-garching.mpg.de/Millennium}}. For merger events that meet the required formation criteria, clusters are created by Monte Carlo sampling at the epoch of the merger and share the metallicity of their host galaxies, with an additional scatter. Both central and satellite halos are followed in the model, and clusters formed within both are collected into the final system. The cluster formation efficiency is linearly proportional to the mass of available cold gas, which in turn is set by the halo mass and redshift. Galaxy metallicity is set by the observed stellar MMR. All the details of the model are described below.

\subsection{Mass Assembly History}

We construct the mass assembly history of dark matter halos using the MM-II simulation \citep{boylankolchin_etal09}. MM-II is a collisionless simulation within a $100 h^{-1}\Mpc$ box, which contains halos up to $\sim 10^{15}\Msun$.  The particle mass, $6.9\times 10^{6}\Msun$, makes it possible to trace the formation of halos as small as $\sim 10^{9}\Msun$.  The cosmological parameters used in the simulation, and adopted in this paper, are $\Omega_{\Lambda}=0.75$, $\Omega_{\mathrm{m}}=0.25$, $h=0.73$, and $\sigma_8=0.9$. 

At first we tried to use the halo catalog of the original Millennium simulation \citep{springel_etal05} but found that its lower mass resolution ($8.6\times 10^{8}\Msun$) does not allow us to track satellite halos less massive than $10^{10}\Msun$.  Since $10^{9}\Msun$ halos can still contribute to forming $10^{5}\Msun$ star clusters that would likely survive to the present day (see Equation~(\ref{eq:mgc})), it is more accurate to use the MM-II catalogs to capture all merger events capable of producing massive star clusters.

The masses of central and satellite progenitors are collected at all 67 outputs from $z=127$ to $z=0$. Parent and child halos are connected with each other in the database by the identifiers \texttt{descendantId} and \texttt{lastProgenitorId}. We apply the tags \texttt{firstProgenitorId} and \texttt{nextProgenitorId} to find the most massive and second most massive progenitors of a given halo, and use their masses to calculate the merger ratio, $R_m$.  This halo merger tree is the starting point of our model.

Although we do not require the halos specifically to be located in a Virgo-sized cluster, it should not bias our comparison with the ACSVCS galaxies.  \citet{cho_etal12} showed that the colors and luminosities of GC systems of early-type galaxies in low-density regions are similar to those in the Virgo cluster.  While the environment has a small effect via the morphology-density relation, the properties of GC systems are primarily dominated by the host galaxy mass.

\subsection{Stellar and Gas Masses}
  \label{sec:baryon}

To all progenitors in a given merger tree, we assign the mass of stars and cold gas according to the following analytical prescriptions.

The stellar mass -- halo mass relation, $M_*(\Mh,z)$, is based on the abundance-matching technique, using a parameterization of \citet[their Equation~(3)]{behroozi_etal13a} for the Sloan Digital Sky Survey (SDSS) measurements of the galaxy luminosity function.

Note that \citet{Kravtsov_etal14} have recently found that the total luminosity, and stellar mass, of central galaxies in clusters (with $\Mh > 10^{14}\Msun$) has been underestimated in the SDSS photometry pipeline, mainly due to over-subtraction of the background light in extended galactic envelopes.  The correction is substantial and can reach a factor of $2-4$ at $M_* \gtrsim 10^{12}\Msun$.  The magnitude of the corresponding correction at $z>0$ is not yet known.  We have decided not to include this correction, because our sample contains only one central cluster galaxy, and more importantly, we use the galaxy stellar mass only as a proxy for estimating the cold gas mass and metallicity from the observed scaling relations, as we describe below.  These relations were derived for the stellar luminosities measured by the SDSS. In order to apply these relations consistently, we use the \citet{behroozi_etal13a} equations as published.

To derive the mass of cold gas, $\Mg$, in a galaxy with stellar mass $M_*$, we combine recent results from the ALFALFA survey \citep{papastergis_etal12} with additional Arecibo observations of nearby starforming galaxies by J.~Bradford \& M.~Geha (in preparation).  These observations measure the mass of neutral HI gas, to which we add the corresponding HeI mass.  We take the measured $\Mg$ as a proxy for the reservoir of gas available for star formation.  The data at $z \approx 0$ show that the mean ratio $\Mg/M_*$ at a given stellar mass exhibits a bend at $M_* \approx 10^{9}\Msun$, and cannot be described by a single power law.  A satisfactory fit is provided by a double power law:
\begin{equation}
 \eta \equiv \frac{\Mg}{M_*} \approx 1.8 \left(\frac{M_*}{10^{9}\Msun}\right)^{-\alpha(M_*)},
 \label{eq:gas_to_stellar}
\end{equation}
with a steeper slope $\alpha=0.68$ for $M_* > 10^{9}\Msun$, and a shallower slope $\alpha=0.19$ for the less massive galaxies with $M_* < 10^{9}\Msun$.  The high-mass slope is consistent with the relation used by MG10 ($\alpha=0.7$), but for dwarf galaxies the gas mass is reduced relative to the MG10 prescription.

The amount of cold gas in high-redshift galaxies is very uncertain.  We can rewrite the gas-to-stellar fraction as $\eta(z) = \sSFR(z)\times \tdep(z)$, where $\sSFR \equiv \SFR/M_*$ is the specific star formation rate (SFR), and $\tdep \equiv \Mg/\SFR$ is the gas depletion timescale. The empirical evolution of the sSFR is consistent with $\sSFR(z)\propto (1+z)^{2.8}$ up to $z \sim 2$ \citep{magdis_etal12a, tacconi_etal13}, while $\tdep(z)$ is consistent with being approximately constant for starforming galaxies at all redshifts \citep[e.g.,][]{bigiel_etal11, feldmann13}. Thus we obtain
\begin{equation}
 \eta(z) = \eta_0 \, (1+z)^{n},
 \label{eq:gas_stars_ev}
\end{equation} 
with $n=2.8$. 

An alternative derivation by \citet{tacconi_etal13}, from a CO survey of molecular gas of actively star-forming galaxies at $z \approx 1-3$, suggests a variable gas depletion timescale, decreasing roughly as $\tdep(z) \approx 1.5\Gyr\, (1+z)^{-1}$, which in turn implies $\eta(z) = \eta_0 \, (1+z)^{1.8}$.  Given the current uncertainty in the gas evolution, when constructing our model we consider both possibilities, $n=2.8$ and $n=1.8$.

At even higher redshift ($z \gtrsim 2-3$), there is evidence that the gas fraction saturates at a maximum value \citep[e.g.,][]{magdis_etal12a}.  Accordingly, we limit $\eta$ for very high redshifts: $\eta(z>3) = \eta(z=3)$.

In addition to the mean relations given by Equations~((\ref{eq:gas_to_stellar}) and (\ref{eq:gas_stars_ev})), we include random scatter of 0.3~dex to account for the combined dispersion of the local MMR, specific SFR, and gas depletion time.

The stellar and gas mass fractions in a given halo are then defined as:
\begin{equation}
  f_* \equiv \frac{M_*}{\fb \, \Mh}, \quad
  f_g \equiv \frac{M_g}{\fb \, \Mh}
  \label{eq:fs}
\end{equation}
where $\fb \approx 0.16$ is the universal baryon fraction \citep[e.g.,][]{hinshaw_etal13}. 

A final constraint of the mass fractions is that the sum of the two cannot exceed the total accreted baryon fraction, $\fin$:
\begin{equation}
  f_*(z) + \fg(z) \leq \fin(z),
\end{equation}
where $\fin \le 1$ is limited by photoheating by the extragalactic UV background, as described in MG10.  In cases when the baryonic fraction ($f_*+f_g$) calculated from Equations~(\ref{eq:gas_to_stellar}) and --(\ref{eq:gas_stars_ev}) exceeds $\fin$, we revise the gas fraction to be $f_{\rm g,revised} \equiv \fin - f_*$.  This constraint only affects halos less massive than $\Mh \sim 10^{10}\Msun$.

\begin{figure}[t]
  \vspace{0cm}
\hspace*{-0.36cm}\includegraphics[width=1.09\hsize]{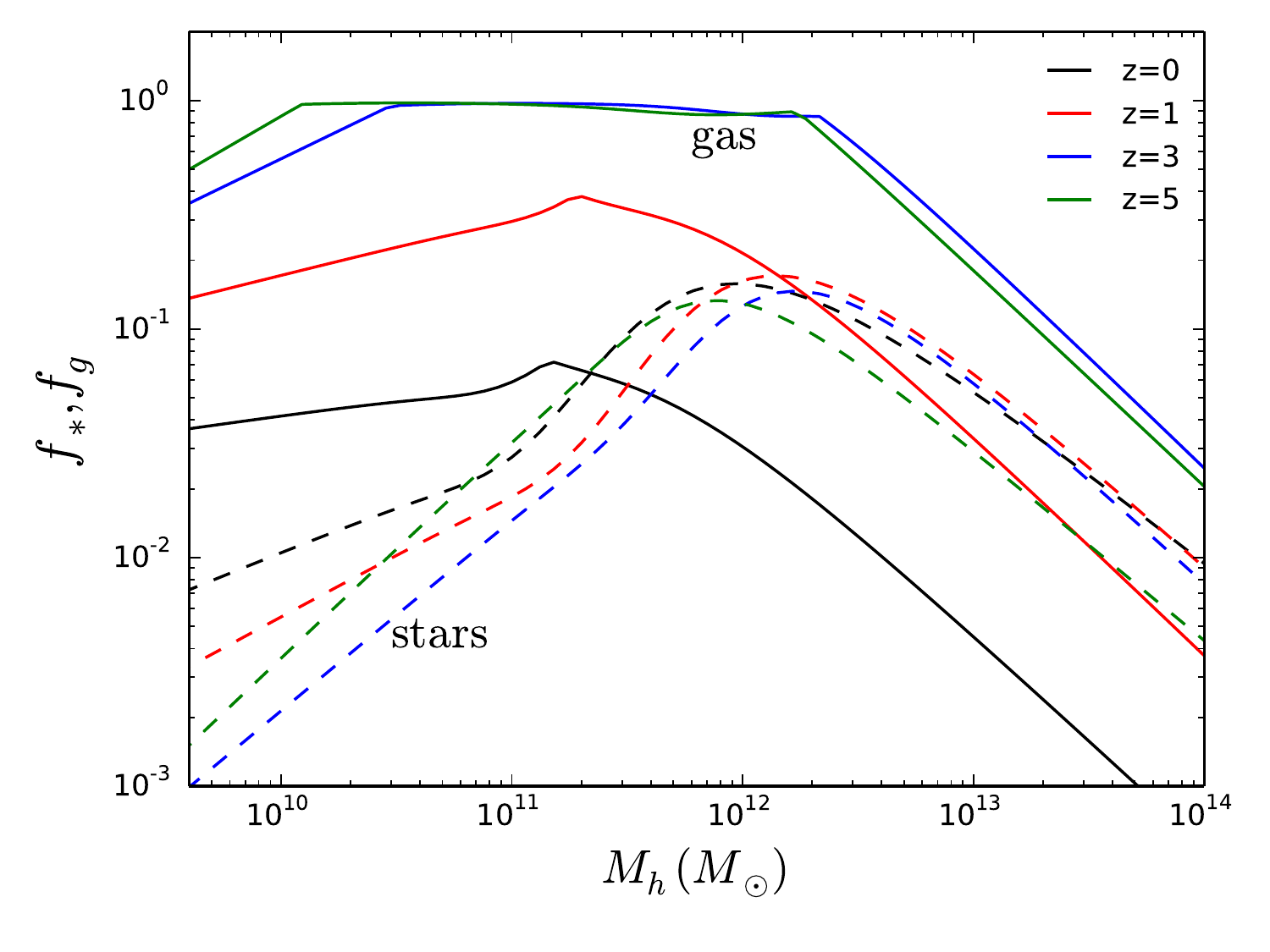}
 \vspace{-0.5cm}
\caption{\small Adopted relation for the fraction of galaxy gas mass (solid lines) and stellar mass (dashed lines), in units of the universal baryon fraction (Equations~(\ref{eq:gas_to_stellar})--(\ref{eq:fs}), with $n=2.8$) vs. halo mass, at several redshifts: $z=0$ (black), $z=1$ (red), $z=3$ (blue), $z=5$ (green).}
  \vspace{0.1cm}
  \label{fig:baryon}
\end{figure}

Figure~\ref{fig:baryon} shows our derived gas and stellar fractions of halos ranging from $10^{9.5}\Msun$ to $10^{14}\Msun$ at redshifts $z=0-5$. The new prescription is similar, but not identical, to that in MG10. The stellar fraction reach its maximum for the Milky Way-sized halos $\Mh\sim 10^{12}\Msun$, and decreases at both higher and lower mass. The new $M_* - \Mh$ relation also depends much less strongly on redshift than that used in MG10.

The evolution of the gas fraction with redshift for the case $n=2.8$ is faster than what is needed to account for the increase of stellar mass of galaxies at $z \lesssim 3$, resulting from the abundance matching method \citep[Figure~4 of][]{behroozi_etal13a}.  We will test the sensitivity of our results to this prescription by considering an alternative calculation of the cold gas mass in Section~\ref{sec:model3}.

\subsection{Cluster Formation}

In our model, clusters are formed during epochs of enhanced star formation following halo mergers. We trace both mergers between a satellite and a central halo, as well as mergers between two satellites. For a halo with mass $M_{\mathrm{h},i}$ at the $i$th simulation output, the mass of its main progenitor and (possible) second largest progenitor at output $i-1$ are $M_{\mathrm{h}, i-1}$ and $M_{\mathrm{h2}, i-1}$, respectively. The merger ratio is defined as $R_m = M_{\mathrm{h2}, i-1}/M_{\mathrm{h}, i-1}$, if the second progenitor is found.  Otherwise, we use the differential increase in halo mass as a proxy: $R_m = (M_{\mathrm{h},i}-M_{\mathrm{h},i-1})/M_{\mathrm{h}, i-1}$. 

A cluster formation event is triggered by a gas-rich major merger, when the merger ratio exceeds a threshold value: $R_m > p_3$. We expect the threshold to be in the range of $p_3 = 0.1-0.5$ to have sufficient influence on the structure of the interstellar medium that could trigger condensation of giant molecular clouds, but the exact value of $p_3$ is an adjustable parameter of the model.

The MG10 model used an additional parameter to set a minimum cold gas fraction of the merging halos (at the level of 4\%).  We have experimented with including this constraint, but found that it is automatically satisfied by the requirement to have enough gas mass to form a cluster with $M > 10^{5}\Msun$, according to Equation~(\ref{eq:mgc}).  Any value of the gas fraction threshold below 10\% gave similar results, and therefore, we set it to zero and eliminate it as a model parameter.

In another departure from the MG10 model, we do not include a ``Case-2'' formation channel here, whereby clusters could form without a detected merger but in extremely gas-rich galaxies with a cold gas fraction above $\approx 98\%$. Instead, in Section~\ref{sec:model4} we will investigate an alternative scenario for continuous cluster formation.

The cluster formation rate scales approximately linearly with the mass of cold gas available for star formation, as indicated by detailed hydrodynamic simulations \citep{kravtsov_gnedin05}:
\begin{equation}
  \MGC = 3\times 10^{-5} p_2 \, \fb^{-1} \, \Mg,
  \label{eq:mgc}
\end{equation}
where $p_2 \sim 1$, the normalization factor, is another adjustable parameter in our model. This relation gives us the total mass of all GCs formed in a given galaxy at a given epoch. The normalization factor $p_2$ is necessary because the galaxy formation cannot be smoothly captured by processing of discrete outputs of the MM-II simulation. Note that the definition of $p_2$ differs from MG10, where it was written as $1+p_2$. Here we explore a wider range of this parameter, allowing for $p_2 < 1$.

The total mass $\MGC$ is then distributed into individual GCs by using a Monte Carlo method and adopting a power-law initial cluster mass function, $dN/dM = M_0 \, M^{-2}$. The minimum mass of individual clusters is set to $M_{\rm min} = 10^{5}\Msun$, below which a typical cluster is expected to be completely evaporated over the Hubble time. After fixing $\MGC$ and $M_{\rm min}$, the maximum cluster mass $M_{\rm max}$ (also equal to the normalization $M_0$) is evaluated from the integral constraint $\MGC = M_{\rm max}\, \ln(M_{\rm max}/M_{\rm min})$.

\begin{figure}[t]
  \vspace{0cm}
\hspace*{-0.35cm}\includegraphics[width=1.1\hsize]{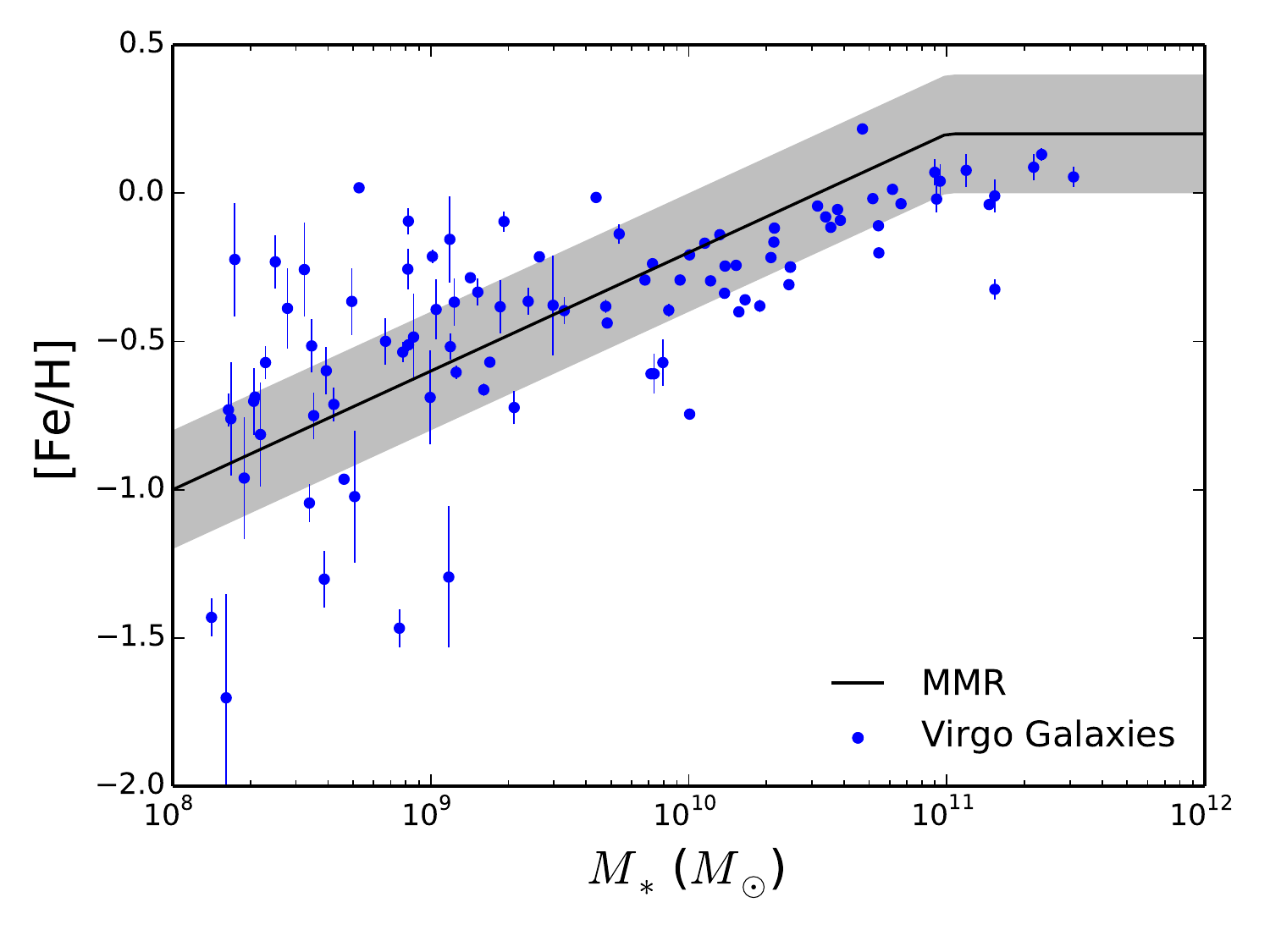}
  \vspace{-0.5cm}
\caption{\small Adopted mass-metallicity relation at redshift zero (solid line) and scatter $\sigmamet = 0.2$~dex (light shading).  Points with errorbars represent the metallicity of the Virgo cluster galaxies, derived from their $(g-z)$ color (see details in Section~\ref{sec:metal}).}
  \vspace{0.1cm}
  \label{fig:MMR}
\end{figure}

\subsection{Mass-Metallicity Relation of Host Galaxies}
  \label{sec:metal}

The metallicity of individual model GC is determined by the metallicity of the host galaxy at the epoch when the GC is created. The stellar MMR of galaxies can be used to estimate the iron abundance of the host.  Based on the observations in the Local Group, the MG10 model adopted a linear relation between the iron abundance and log stellar mass at zero redshift: $\feh = 0.4 \, \log(M_*/10^{10.5}\Msun)$, along with a gradual evolution of this relation with redshift. 

Recent observational evidence suggests that the MMR is better described as a two-dimensional projection of a three-dimensional fundamental metallicity relation (FMR) of stellar mass, metallicity, and SFR \citep[e.g.][]{laralopez_etal10, laralopez_etal13}, or alternatively, HI gas mass \citep{bothwell_etal13}. No additional redshift evolution is needed.  By combining this FMR with the evolution of the gas-to-stellar mass ratio, we can derive an explicit redshift-dependence of the projected MMR. As suggested by \citet{bothwell_etal13}, a new variable ($\log M_* - 0.35 \log \Mg$) minimizes the scatter in the FMR. In the notation of our Equation~(\ref{eq:gas_stars_ev}) this variable can be rewritten as ($\log M_* - 0.54 \, n \log(1+z)$). Substituting it for $\log M_*$ in the local MMR, we obtain the evolving relation:
\begin{equation}
  \feh = 0.4 \, \log\left(\frac{M_*}{10^{10.5}\Msun}\right) - 0.216 \, n\log(1+z).
  \label{eq:metal_evo}
\end{equation}
The power-law slope, and amount of evolution to $z\approx 0.7$, is consistent with the recent measurements of \citet{gallazzi_etal14}. The redshift-dependence of our MMR is also consistent with the observed evolution of the gas-phase (O/H) abundance in the AGES survey \citep{moustakas_etal11} and 3D-HST survey \citep{cullen_etal14}: about 0.3~dex lower metallicity at $z \approx 2$ relative to $z=0$, at a fixed stellar mass. \citet{zahid_etal14} derived somewhat different slopes of the mass and redshift dependence using the DEEP2 and COSMOS data, but the average amount of evolution to $z\approx 1.6$ is the same as in our relation, 0.25~dex.

It should be noted that the linear scaling with log stellar mass is not valid for very massive galaxies, whose metallicity tends to saturate at a supersolar value. Accordingly, in our model we limit the galaxy metallicity to the maximum value $\feh = 0.2$ for $M_* > 10^{11}\Msun$.

Figure~\ref{fig:MMR} shows our adopted MMR, along with derived metallicities of the Virgo cluster galaxies. The latter are calculated from the luminosity-weighted $(g-z)$ color using an empirical color-metallicity relation obtained by \citet{peng_etal06}, Equation~(\ref{eq:Peng06}) below, from the Galactic GC data.  Applying this relation to the Virgo galaxies containing a mixture of stellar populations is justified because the resulting metallicities are clustered around the observed galactic MMR, without systematic bias. The stellar mass of these galaxies is converted from the K-band luminosity and the color-dependent mass-to-light ratio from \citet{bell_etal03}.

\citet{kirby_etal13} showed that nearby dIrr galaxies follow the same mean MMR as dSph galaxies, with the power-law slope $\approx 0.3$. The normalization of their MMR matches our MMR at $M_* \approx 10^{8}\Msun$, but falls a little lower for higher mass galaxies.  Given the considerable dispersion in the derived metallicities of the local and Virgo cluster galaxies, the two relations are not necessarily inconsistent.

\begin{figure}[t]
  \vspace{0cm}
\hspace*{-0.35cm}\includegraphics[width=1.05\hsize]{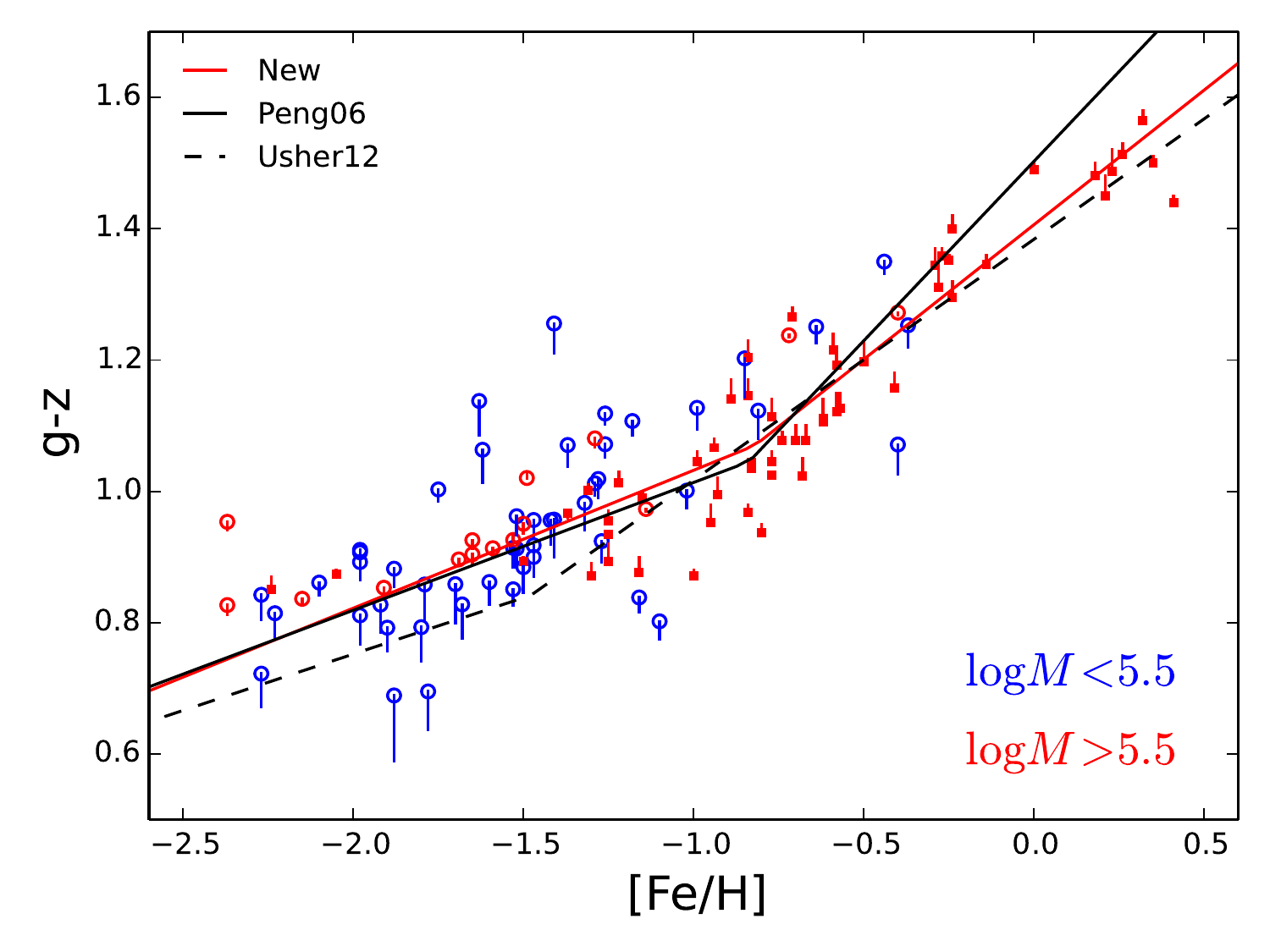}
 \vspace{-0.3cm}
\caption{\small Color-metallicity relation of Galactic (open circles) and extragalactic (squares) GCs with spectroscopically measured [Fe/H].  Symbols show the $(g-z)$ color corrected for the evolutionary mass dependence, described by Equation~(\ref{eq:color_corr}) and corresponding text in Section~\ref{sec:color_metal}.  Vertical lines leading to the symbols show the amount and direction of this correction. Our new color-metallicity relation is overplotted (red line; Equation~(\ref{eq:color_metal_new})) together with the other relations derived by \citet{peng_etal06} (black solid line) and \citet{usher_etal12} (black dashed line).}
  \vspace{0cm}
  \label{fig:color_metal}
\end{figure}

\subsection{Revisiting the GC Color-Metallicity Relation}
  \label{sec:color_metal}

The metallicity distribution of our model GCs should be compared with the observed populations. Unfortunately, spectroscopic measurements of GC metallicity outside the Local Group are rare. A common approach for Virgo cluster galaxies is to convert the observed GC colors via an empirical color-metallicity relation. \citet{peng_etal06} derived such an empirical relation based on the calibration with the Galactic GCs and the additional clusters in Virgo galaxies M49 and M87 with available spectroscopy. The relation is nonlinear, with a steeper slope for metal-poor clusters ($\feh < -0.8$):
\begin{eqnarray}
\feh & = & -6.21 + 5.14 (g-z), \; \rm if \; 0.70<(g-z)<1.05 \nonumber\\
\feh & = & -2.75 + 1.83 (g-z), \; \rm if \; 1.05<(g-z)<1.45
  \label{eq:Peng06}
\end{eqnarray}
\citet{usher_etal12} used calcium triplet-based spectroscopy to determine the metallicity of 903 GCs in 11 early-type galaxies. They found a similarly nonlinear color-metallicity relation but with a different break point and slopes. Most recently, \citet{vanderbeke_etal14} obtained updated SDSS photometry of 96 Galactic GCs and suggested a cubic polynomial fit for the relation.

However, based on the models for GC evolution, \citet{goudfrooij_kruijssen14} point out that as clusters lose preferentially low-mass (red) stars by evaporation, their color tends to get bluer. The process is driven by two-body relaxation, which is faster in lower mass clusters. This introduces a possible mass-dependent bias in the color-metallicity relation. To test for this bias, we took the data from \citet{vanderbeke_etal14} for 96 Galactic GCs with the metallicity from the updated \citet{harris96} catalog, and added the clusters in M49 and M87 from \citet{peng_etal06}. The mass of an individual GC can be estimated by using the color-dependent mass-to-light ratio from \citet{bell_etal03}: $\log(M/L_z) = 0.322 \, (g-z) - 0.171$. We fit a linear model of two variables, $(g-z) = \beta_0 + \beta_1 \, \feh +\beta_2 \, \log{M}$, and test whether the mass dependence is significant based on the ANOVA variance method. The $p$-value of the test statistic is $\lesssim 10^{-5}$, which means the null hypothesis that the color is independent of mass can be safely rejected. Thus the available data support the theoretical expectation of \citet{goudfrooij_kruijssen14} that low-mass GCs develop a "blue-shift" relative to high-mass GCs.  This evolutionary change of color is in addition to the gradual reddening due to the passive stellar evolution, which is independent of cluster mass.

In order to use the color-metallicity relation to infer $\feh$ of the observed clusters, we need to "undo" this evolutionary effect.  We introduce a simple correction to the color
\begin{equation}
 (g-z)_{\mathrm{cor}} = (g-z)_0 - 0.03 \, \log{\left(M \over 10^{6}\Msun\right)}
  \label{eq:color_corr}
\end{equation} 
that minimizes the mass-dependence of the color-metallicity relation. Then we refit the nonlinear relation and obtain
\begin{eqnarray}
 (g-z)_{\mathrm{cor}} & = & 0.21 (\feh+0.82)-1.07, \; \rm if \; \feh<-0.82 \nonumber\\
 (g-z)_{\mathrm{cor}} & = & 0.41 (\feh+0.82)-1.07, \; \rm if \; \feh\geq -0.82
 \label{eq:color_metal_new}
\end{eqnarray}
shown in Figure~\ref{fig:color_metal}. We use this relation to determine GC metallicities in the Virgo galaxies and compare them to our model.

\section{Dynamical Disruption}
  \label{sec:dyn}
  
Although GCs are relatively stable and long-lived self-gravitating systems, they still gradually lose stars and dissolve into the field. In this paper, we include two sources of the mass loss: the evaporation of stars via two-body relaxation, which reduces the number of stars $N_*$ within the cluster, and stellar winds and explosions, which reduce the average stellar mass $\bar{m}$. These two mechanisms can be quantified by the mass continuity equation:
\begin{equation}
\frac{1}{M}\frac{dM}{dt}\equiv\frac{1}{N_*}\frac{dN_*}{dt}+\frac{1}{\bar{m}}\frac{d\bar{m}}{dt}=-\nu_{\rm ev}(M)-\nu_{\rm se}\frac{\bar{m}(0)}{\bar{m}},
  \label{eq:mass-con}
\end{equation}
where $\nu_{\rm ev}$ and $\nu_{\rm se}$ are the evaporation rate and mass-loss rate due to stellar evolution, respectively. The time-dependent mass-loss rate $\nu_{\rm se}$ for a \citet{kroupa01} IMF is derived in \citet{prieto_gnedin08}. The cluster evaporation rate is estimated via the half-mass relaxation time:
\begin{equation}
  \nu_{\rm ev} = \frac{\xi_e}{t_{\rm rh}}=\frac{7.25 \, \xi_e \, \bar{m} \, G^{1/2} \, \ln\Lambda}{M^{1/2} \, R_{\mathrm{h}}^{3/2}},
  \label{eq:evaporation}  
\end{equation}
where $\xi_e=0.033$ is the escape fraction of stars per relaxation time, $R_{\mathrm{h}}$ is the half-mass radius, $\bar{m}=0.87\Msun$ is the average stellar mass for a Kroupa IMF, and $\ln\Lambda=12$ \citep[e.g.,][]{spitzer87}. 

By assuming that the stellar evolution mass-loss timescale is much shorter than the evaporation timescale, the decrease of the initial cluster mass with time can be calculated as
\begin{equation}
 M(t) = M(0) \left[1-\int^t_0\nu_{\rm se}(t')dt'\right] \left[1-\frac{1+3\delta}{2}\nu_{\rm ev,0} \, t\right]^{2/(1+3\delta)},
  \label{eq:dyn}
\end{equation}
where we take $\delta = 2/3$, as in MG10. The above parameterization provides a good fit to the results of direct N-body simulations of tidally limited clusters.

\section{Alternative Models}
  \label{sec:other_models}

Thus far, we have constructed a model for cluster formation during gas-rich merger events. We will refer to the above prescription as Model 1. There are three major uncertainties in this model: the evolution of the MMR with cosmic time (Equation~(\ref{eq:metal_evo})), the evolution of the cold gas fraction (Equation~(\ref{eq:gas_stars_ev})), and the need for major mergers to trigger cluster formation.

\subsection{Model 2: No Metallicity Evolution}

Although there is a consensus that more massive galaxies have higher metallicity, recent studies suggest that the MMR is more complex than was expected before. As summarized in Section~\ref{sec:metal}, there exists a fundamental plane for star-forming galaxies in the three-dimensional parameter space: SFR, metallicity, and stellar mass. On the other hand, scarcity of spectroscopic observations of high-redshift galaxies leaves the evolution of the MMR with time very uncertain. In order to test the sensitivity of our model to this relation, we consider an extreme case of no-evolution and apply the local MMR, $\feh=0.4\log(M_*/10^{10.5}\Msun)$, at all redshifts. We keep the rest of the prescriptions as in Model 1, and refer to this new case as Model 2.

\subsection{Model 3: Gas Mass from SFR}
  \label{sec:model3}
  
In Model 1, the gas mass-stellar mass relation is used to derive the cold gas fraction of the halos. An alternative way of determining the gas mass is to use the gas depletion timescale, defined as the ratio between the cold gas mass and SFR, $\tdep \equiv \Mg/\SFR$. \citet{bigiel_etal11} found a constant timescale $\tau_{\rm Dep} \approx 2.35\Gyr$, with $1\sigma$ scatter of 0.24~dex. This depletion time, together with the SFR required to match the growth of the stellar mass of galaxies \citep{behroozi_etal13b}, can be used to determine the amount of gas available for star formation. Note that such an empirical derivation of the SFR is independent of the direct measurements in Lyman-break galaxies, discussed in Section~\ref{sec:baryon}. In particular, it leads to much lower gas fraction in halos with $\Mh < 10^{11}\Msun$ at redshift $z>2$, relative to that in Model 1. These halos are crucial for cluster formation, as they are the typical hosts of the metal-poor GC population. Observational surveys of low-mass high-redshift galaxies are greatly incomplete and may be underestimating the stellar mass growth. To correct for the possible incompleteness, we set the gas fraction in halos with $\Mh < 10^{11}\Msun$ at $z>2$ to be the same as in a $10^{11}\Msun$ halo at $z=2$. We keep the rest of the prescriptions as in Model 1, and refer to this new case as Model 3.

\subsection{Model 4: No Mergers}
  \label{sec:model4}
  
Since all of the models above are based on the assumption that GCs are formed in gas-rich mergers, one may ask whether the merger scenario is a required channel for cluster formation or just one of several possible ways to reproduce the GC bimodality. Another plausible formation channel is during galactic starbursts, characterized by enhanced SFR, regardless of whether they are caused by mergers or continuous gas accretion. In this starburst case, the trigger for cluster formation can be either an SFR or specific SFR exceeding a critical threshold, $\SFR_{\mathrm{crit}}$ or $\sSFR_{\mathrm{crit}}$. We calculate the sSFR (SFR) for the halos in the whole merger tree from the differential stellar mass growth, as described in \citet{behroozi_etal13b}, and trigger cluster formation when the sSFR (SFR) exceeds the threshold.  The latter is a new parameter of this alternative model, to which we refer as Model 4.

\begin{figure}[t]
  \vspace{0cm}
\hspace*{-0.35cm}\includegraphics[width=1.1\hsize]{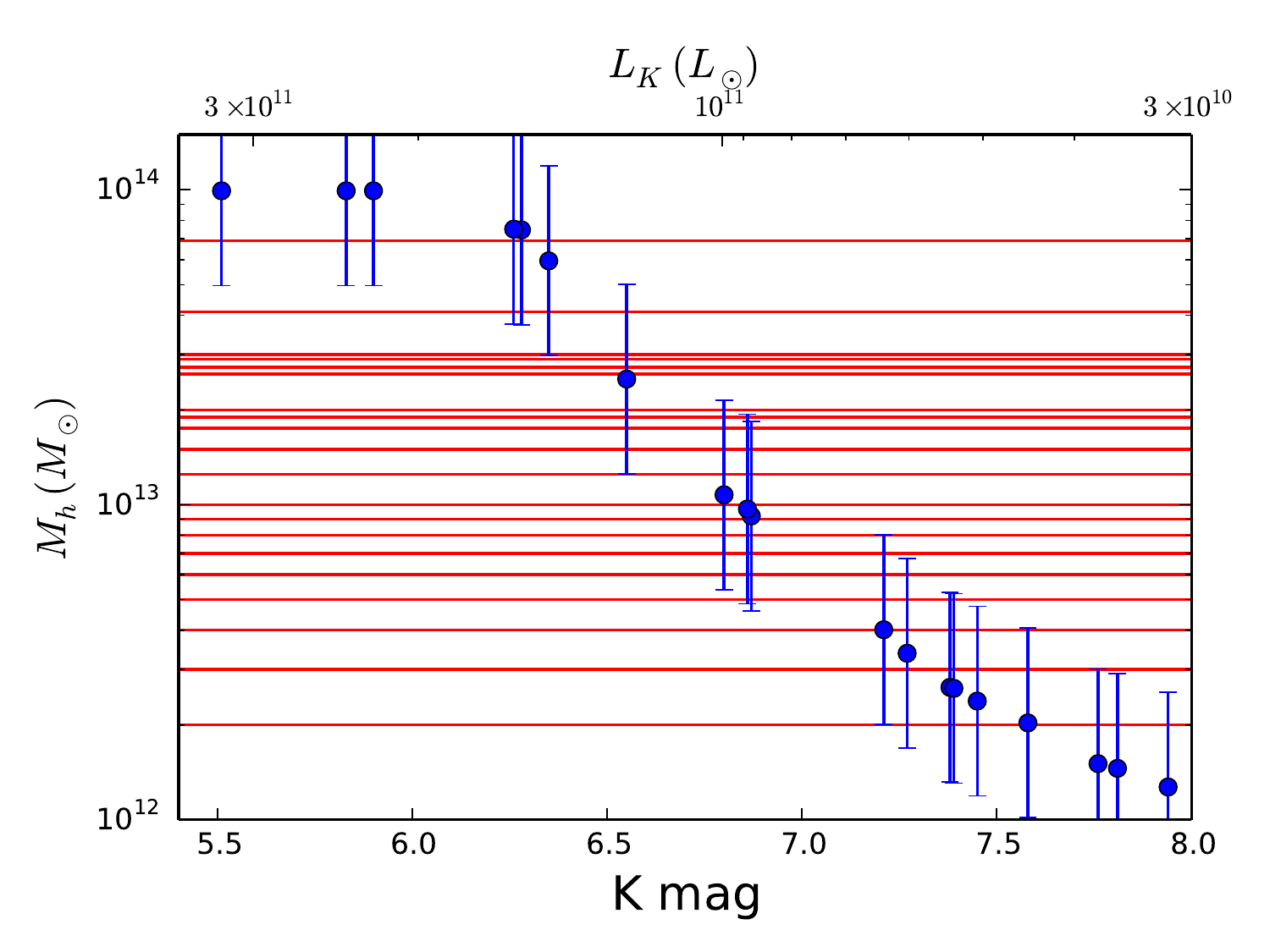}
  \vspace{-0.4cm}
\caption{\small K-band apparent magnitude vs. derived halo mass for 19 Virgo galaxies, using the abundance matching technique (see Section~\ref{sec:matching}). Red horizontal lines show the masses of the 20 halos selected from the MM-II simulation.}
  \vspace{0.2cm}
  \label{fig:pair}
\end{figure}

\section{Results}
  \label{sec:result}
  
\subsection{Galaxy-Halo Matching}
  \label{sec:matching}

The above prescriptions allow us to create GCs within the MM-II halos with different mass assembly histories. We then compare the metallicity distribution of the model GC population to the observed GC systems in the Virgo cluster galaxies. In order to match the galaxies to the halos, we take the stellar mass of the Virgo galaxies, obtained from their color and K-band magnitude (Section~\ref{sec:metal}), and calculate the expected mass of their halo using the stellar mass-halo mass relation \citep{behroozi_etal13a}. As discussed in Section~\ref{sec:baryon}, this relation may be based on an underestimated stellar mass of giant galaxies, which would lead to an overestimate of derived halo mass. To correct for this, we set the maximum halo mass at $\sim 10^{14}\Msun$. The photometry of the Virgo galaxies, such as K-band magnitude and color, are from the 2MASS catalog provided on the ACSVCS Website\footnote{\url{https://www.astrosci.ca/users/VCSFCS/Home.html}}.  The effective radii $R_{e}$ are obtained by fitting the Sersic profile \citep{ferrarese_etal06}. These data are reproduced in Table~\ref{tab:host_galaxy}, along with the derived stellar and halo mass. The GC Virgo catalogs with the SDSS $g$ and $z$-band magnitudes are from \citet{jordan_etal09}.

Because of the uncertainty in both the mass-to-light ratio and $M_*-\Mh$ relation, we add random scatter (0.1~dex or 0.2~dex) for each conversion step. The number of matching galaxy-halo pairs increases with increasing scatter, but the best-fit model parameters are not sensitive to the exact value. The relation between the observed K-band magnitude and calculated halo mass for the Virgo galaxies is plotted in Figure~\ref{fig:pair}, together with the MM-II halos. We selected only halos with $\Mh \ga 10^{12}\Msun$ to model massive elliptical galaxies that contain largest samples of clusters. 
%\textbf{Some basic properties of the Virgo galaxies which matched with our 20 selected halos are listed in Table(\ref{tab:host_galaxy}).}

The galaxy-halo matching procedure is straightforward. We match each galaxy to all MM-II halos that fall within its calculated mass range, shown by the error bars in Figure~\ref{fig:pair} and listed in Table~\ref{tab:host_galaxy}. For each pair, we compare the model and observed GC metallicity distributions using the Kolmogorov-Smirnov (KS) test. Then we combine the $p$-values of the KS probability for all pairs into a joint set and calculate the fraction of pairs with the $p$-value larger than 0.01. This fraction defines the "goodness" of our model, $G_{0.01}$. The 1\% level of the KS probability is generally considered to indicate that the model is not inconsistent with the data. A "goodness" value of $G_{0.01} = 50\%$ means that half of the model realizations are consistent with the observed GC metallicities. The best-fit model parameters are determined by maximizing the "goodness" value.

\begin{table}
\begin{center}
\caption{\sc Host Galaxy Properties}
  \label{tab:host_galaxy}
\begin{tabular}{rrrrcc}
\tableline\tableline\\
VCC ID &  K mag & $(g-z)$ &  $R_{e}\,(\mathrm{kpc})$ & $M_* \,(10^{10}\Msun)$ & $\Mh \,(10^{12}\Msun)$
\\[2mm] \tableline\\
 1226 &   5.51 &    1.60 &   17.0  & 31.0 & $50-198$ \\
 1316 &   5.90 &    1.60 &   13.7  & 21.7 & $50-198$ \\
 1978 &   5.83 &    1.62 &    8.0  & 23.3 & $50-198$ \\
  881 &   6.28 &    1.57 &   35.3  & 15.3 & $37-149$ \\
  798 &   6.26 &    1.38 &   12.9  & 15.4 & $38-150$ \\
  763 &   6.35 &    1.56 &   12.7  & 14.6 & $30-119$ \\
  731 &   6.80 &    1.53 &    9.9  &  9.4 & $5.4-21$ \\
 1535 &   6.55 &    1.59 &   10.0  & 11.9 & $12.6-50$ \\
 1903 &   6.87 &    1.53 &   10.1  &  9.0 & $4.6-18$ \\
 1632 &   6.86 &    1.61 &    6.9  &  9.1 & $4.9-19$ \\
 1231 &   7.27 &    1.53 &    1.5  &  6.2 & $1.7-6.7$ \\
 2095 &   7.45 &    1.44 &    1.1  &  5.2 & $1.2-4.8$ \\
 1154 &   7.21 &    1.54 &    2.4  &  6.6 & $2.0-8.0$ \\
 1062 &   7.38 &    1.53 &    1.1  &  5.5 & $1.3-5.3$ \\
 2092 &   7.58 &    1.50 &    2.5  &  4.7 & $1.0-4.1$ \\
  369 &   7.94 &    1.57 &    0.6  &  3.4 & $0.64-2.5$ \\
  759 &   7.81 &    1.54 &    2.2  &  3.8 & $0.73-2.9$ \\
 1692 &   7.76 &    1.53 &    0.8  &  3.9 & $0.76-3.0$ \\
 1030 &   7.39 &    1.49 &    0.8  &  5.4 & $1.3-5.2$
\\[2mm] \tableline
\end{tabular}
\end{center}
\vspace{0.1cm}
\end{table}

%\begin{table}
%\begin{center}
%\caption{\sc Host Galaxy properties}
%  \label{tab:host_galaxy}
%\begin{tabular}{rrrrrr}
%\tableline\tableline\\
%VCC ID &  K mag & $(g-z)$ &  $R_{e}\,(\mathrm{kpc})$ & $M_* \,(10^{10}\Msun)$ & $\Mh \,(10^{12}\Msun)$
%\\[2mm] \tableline\\
% 1226 &   5.51 &    1.6  &   17    & 30.98 & [49.78, 198.12] \\
% 1316 &   5.9  &    1.6  &   13.66 & 21.7  & [49.78, 198.12] \\
% 1978 &   5.83 &    1.62 &    8    & 23.3  & [49.78, 198.12] \\
%  881 &   6.28 &    1.57 &   35.28 & 15.33 & [37.41, 148.88] \\
%  798 &   6.26 &    1.38 &   12.85 & 15.35 & [37.61, 149.68] \\
%  763 &   6.35 &    1.56 &   12.65 & 14.59 & [29.83, 118.72] \\
%  731 &   6.8  &    1.53 &    9.93 &  9.42 & [ 5.39,  21.47] \\
% 1535 &   6.55 &    1.59 &   10    & 11.89 & [12.57,  50.04] \\
% 1903 &   6.87 &    1.53 &   10.12 &  8.98 & [ 4.62,  18.39] \\
% 1632 &   6.86 &    1.61 &    6.88 &  9.13 & [ 4.86,  19.35] \\
% 1231 &   7.27 &    1.53 &    1.49 &  6.16 & [ 1.69,   6.74] \\
% 2095 &   7.45 &    1.44 &    1.14 &  5.17 & [ 1.19,   4.75] \\
% 1154 &   7.21 &    1.54 &    2.36 &  6.64 & [ 2.01,   8.01] \\
% 1062 &   7.38 &    1.53 &    1.14 &  5.45 & [ 1.32,   5.25] \\
% 2092 &   7.58 &    1.5  &    2.52 &  4.71 & [ 1.02,   4.05] \\
%  369 &   7.94 &    1.57 &    0.62 &  3.39 & [ 0.64,   2.54] \\
%  759 &   7.81 &    1.54 &    2.18 &  3.77 & [ 0.73,   2.91] \\
% 1692 &   7.76 &    1.53 &    0.75 &  3.87 & [ 0.76,   3.01] \\
% 1030 &   7.39 &    1.49 &    0.75 &  5.43 & [ 1.31,   5.21] \\
%\\[2mm] \tableline
%\end{tabular}
%\end{center}
%\vspace{0.1cm}
%\end{table}

\begin{table}
\begin{center}
\caption{\sc Fiducial Parameters of Model 1}
  \label{tab:par}
\begin{tabular}{llcl}
\tableline\tableline\\
\multicolumn{1}{l}{Parameter} &
\multicolumn{1}{l}{Best} &
\multicolumn{1}{c}{Range} &
\multicolumn{1}{l}{Effect}\\
\multicolumn{1}{l}{} &
\multicolumn{1}{l}{value} &
\multicolumn{1}{c}{considered} &
\multicolumn{1}{l}{}
\\[2mm] \tableline\\
$\sigmamet$ & 0.2  & $0.1-0.2$ & Scatter of MMR\\
$p_2$       & 2.6  & $1-5$     & Normalization of cluster formation rate\\
$p_3$       & 0.33 & $0-0.5$   & Minimum merger ratio\\
$n$		    & 2.8  & $1.8-2.8$ & Index of cold gas evolution
\\[2mm] \tableline
\end{tabular}
\end{center}
\vspace{0.1cm}
\end{table}

\begin{figure}[t]  
  \vspace{0cm}
\hspace*{-0.35cm}\includegraphics[width=1.05\hsize]{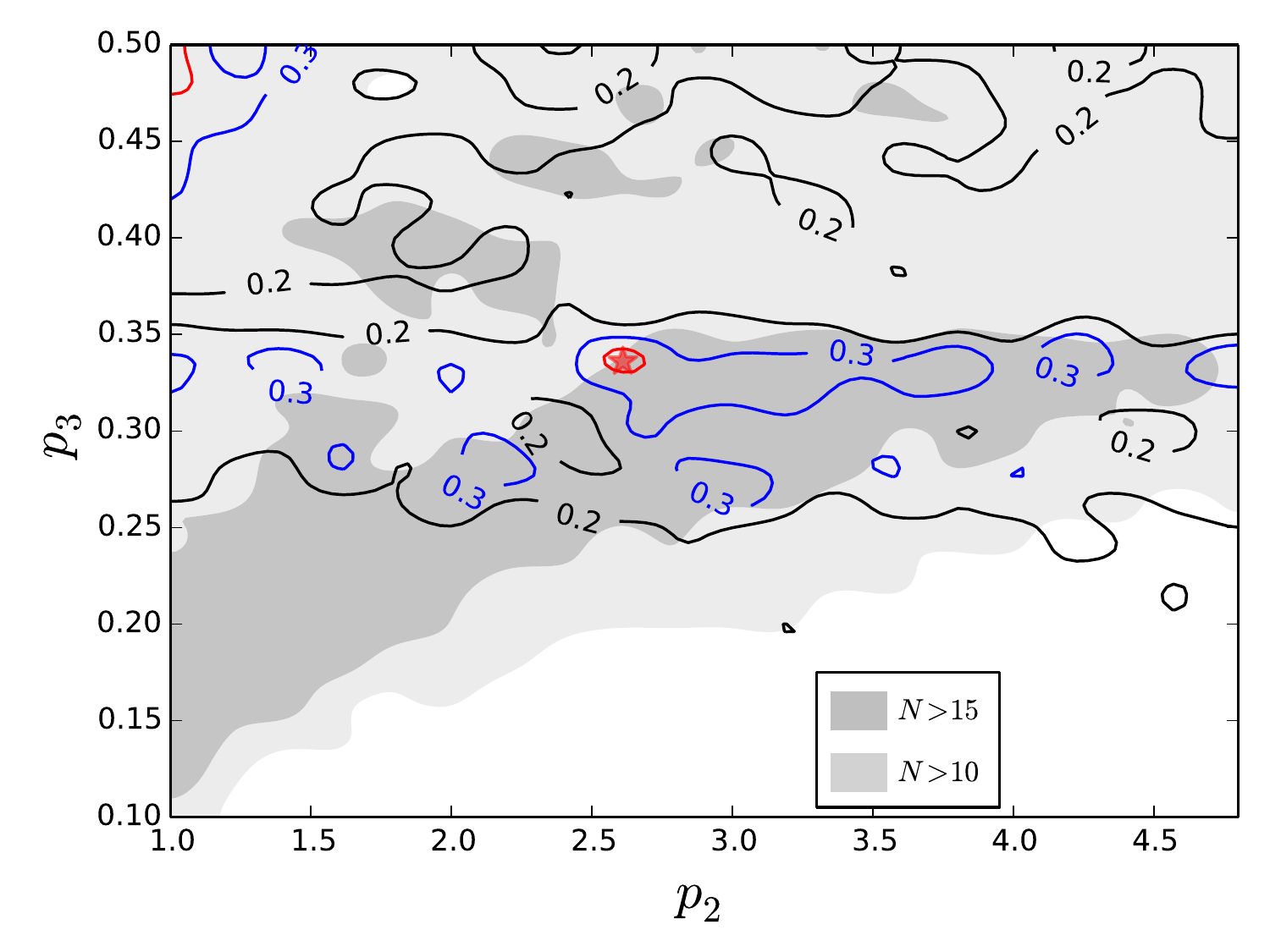}
  \vspace{-0.3cm}
\caption{\small "Goodness" contours on the $p_2-p_3$ parameter plane for Model 1, with fixed $\sigma_{\rm met}=0.2$ and $n=2.8$. For example, a contour marked with "0.3" encloses the range of parameters with $G_{0.01} > 30\%$.  Shaded regions show the number of galaxies with the size of their GC system sufficiently similar to the observed (see Section~\ref{sec:explore} for details). The fiducial model with the best-fit parameters $p_2=2.6$ and $p_3=0.33$ is labeled by a red star.  This model has both the highest $G_{0.01}$ value and the largest number of sufficient GC systems.}
  \vspace{0cm}
  \label{fig:p2p3_model1}
\end{figure}

\begin{figure*}[t]
  \vspace{-0.3cm}
\hspace*{-1.1cm}\includegraphics[width=1.15\hsize]{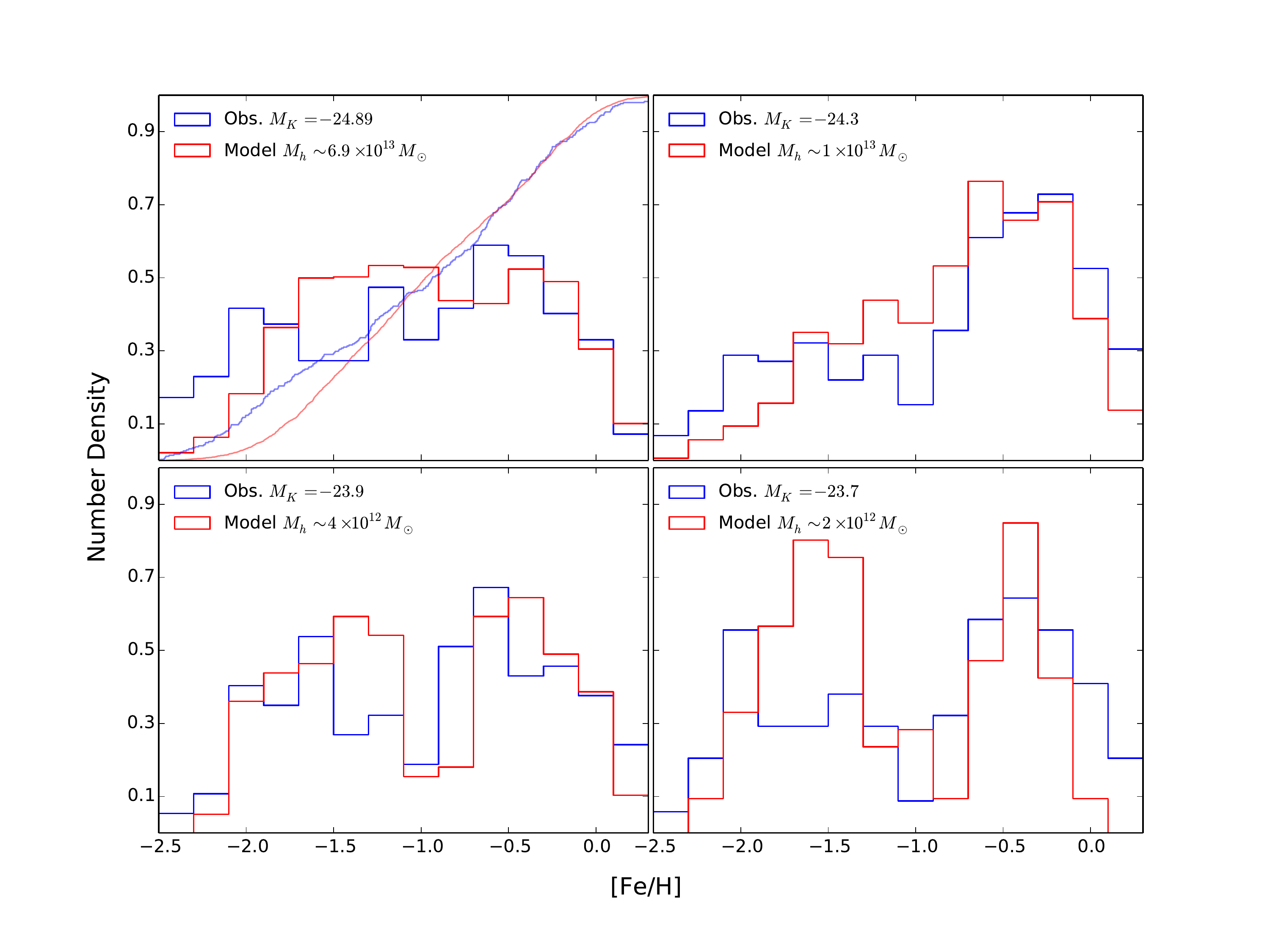}
  \vspace{-1cm}
\caption{\small Comparison between the observed and modeled GC metallicity distributions. Different panels represent the halos of different mass, from largest to smallest, and their matched Virgo cluster galaxies. In the upper left panel, we overplot the cumulative distributions to show that the model is consistent with the data.  The KS test probabilities for the four panels are $p_{KS} \approx 1\%, 4\%, 49\%, 7\%$, in order of decreasing mass.}
  \vspace{0.4cm}
  \label{fig:four-samples}
\end{figure*}

\subsection{Exploration of the Parameter Space}
  \label{sec:explore}

The adjustable model parameters are summarized in Table~\ref{tab:par}. In order to explore the space of parameters $p_2$ and $p_3$, we calculate $G_{0.01}$ on a two-dimensional grid spanning the range of $1 \le p_2 \le 5$, $0 < p_3 \le 0.5$. Smaller values of the formation rate factor, $p_2 < 1$, lead to an insufficient amount of gas to form enough clusters to match the observations. At first, we fix the other two parameters, $\sigmamet=0.2$ and $n=2.8$, and investigate them in detail later.

Figure~\ref{fig:p2p3_model1} shows the contours of $G_{0.01}$, up to the maximum value of about 40\%.  This is a significant enough fraction of galaxy-halo pairs with the model GC metallicities matching the observations.

In addition to the metallicity distribution, an important statistic is the total number of clusters surviving dynamical disruption to redshift zero, that is, the size of the current GC system. It would be very unlikely for any model to produce exactly the observed number of clusters in a given galaxy. Therefore, we introduce a "tolerance" of 0.2~dex on the logarithm of the ratio of the number of model clusters to observed clusters.  If $|\log(N_{\rm GC,model}/N_{\rm GC,obs})| < 0.2$, we consider it a "sufficient" GC system.  Shaded regions in Figure~\ref{fig:p2p3_model1} show the number of galaxies with the sufficient GC systems.  The darkest shade indicates an almost complete match: 16, 17, 18, or 19 systems for our total sample of 19.

The region with the highest numbers of sufficient systems lies near the line pointing from $(p_2, p_3 = 1.0, 0.2)$ to $(4.5, 0.35)$. This trend can be easily understood: the larger the boosting factor $p_2$, the more GCs are created in the model, which need to be compensated by fewer merger events, and therefore, larger threshold ratio $p_3$.

The number of sufficient systems $N_s$ helps us select the parameters of the best-fit fiducial model. There are two sets of parameters with equally high goodness $G_{0.01} \approx 40\%$: $(p_2, p_3 = 2.6, 0.33)$ marked by a red star, and $(1.0, 0.48)$.  However, the second set has significantly lower $N_s$, and consequently, we discard it.

We have also varied the scatter of the MMR $\sigmamet = 0.1$ \& 0.2, and found that $\sigmamet=0.2$ gives a higher goodness value, simply because it can spread the metallicity range to reach the low ($\feh<-2$) and high ($\feh>0$) tails of the observed distribution.

The index of the cold gas fraction $n$ (Equation~(\ref{eq:gas_stars_ev})) is suggested to have two values: 1.8 and 2.8, as we discussed in Section~\ref{sec:baryon}. Intuitively, $n=1.8$ will lead to a slower increase of $\fg$ toward high redshift, which would suppress the formation of GCs. It will also slow down the metallicity evolution, and in turn bring closer the red and blue peaks of the $\feh$ distribution. We have repeated the model-selection procedure for the $n=1.8$ case and explored the goodness contours in the $p_2-p_3$ plane, for different $\sigmamet$. The largest goodness value is only $G_{0.01} = 0.16$, much smaller than that in the $n=2.8$ case. Thus we conclude that $n=2.8$ is favored in our model.

\begin{figure}[t]  
  \vspace{0cm}
\hspace*{-0.35cm}\includegraphics[width=1.05\hsize]{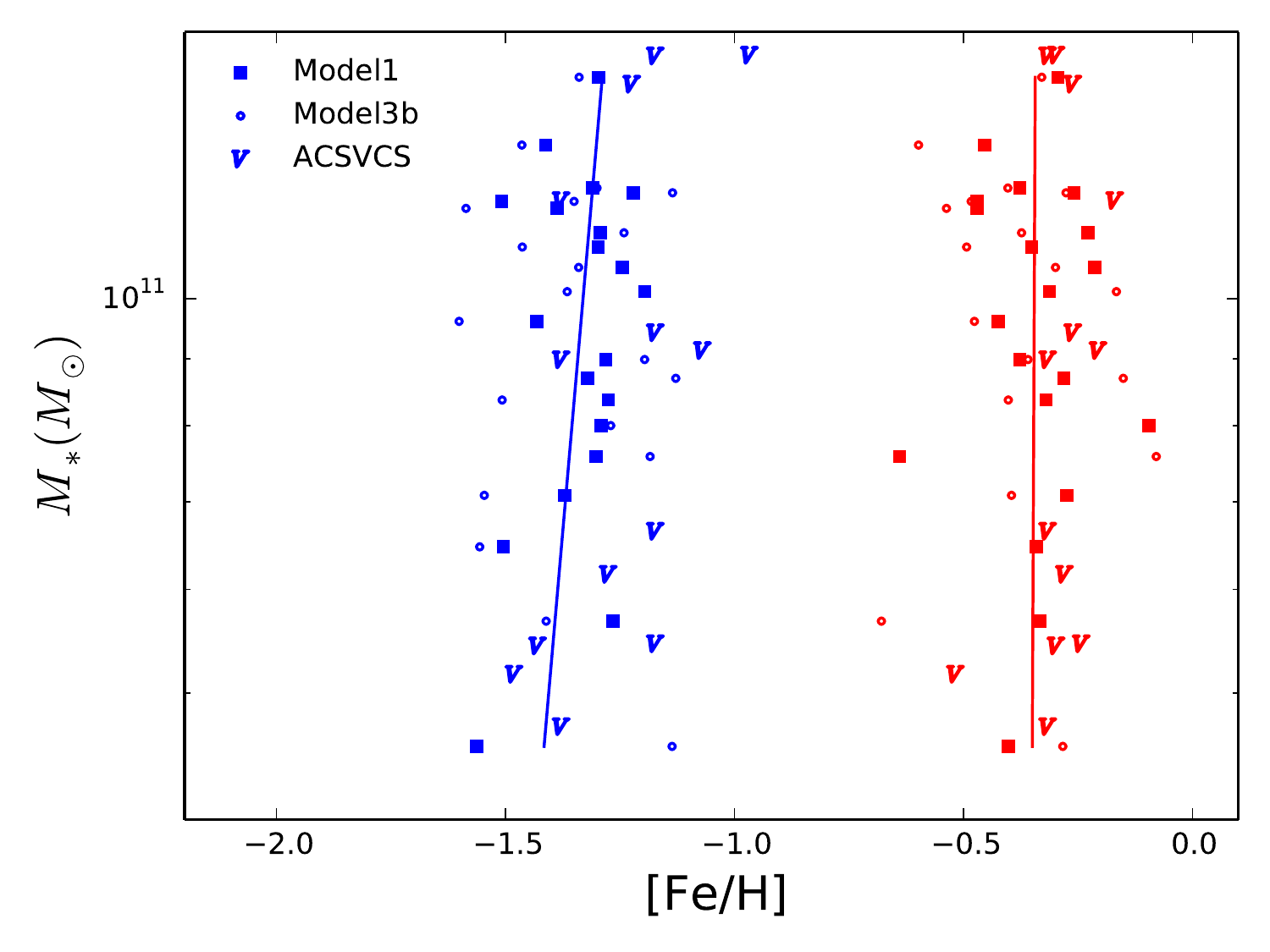}
  \vspace{-0.3cm}
\caption{\small Peak metallicities of the blue and red cluster subpopulations within the Virgo galaxies (symbols V) and the fiducial Model 1 (squares) and Model 3 (circles) halos. Solid lines show linear fit to Model 1 points.}
  \vspace{0.1cm}
  \label{fig:peaks}
\end{figure}

\subsection{Removal of GCs of Satellite Galaxies} \label{sec:removal_sat}

A caveat to our comparison is that the observed samples cover only inner parts of the Virgo galaxies. The ACSVCS data were obtained from single pointings of the HST/ACS camera with the field of view $202\arcsec \times 202\arcsec$, which corresponds to $\sim 16\times16\kpc$ at the distance of the Virgo cluster. For 8 of our 19 galaxies, this scale lies within the effective diameter of the stellar distribution and therefore many GCs may be located outside. \citet{peng_etal08} extrapolated the GC number density profiles to larger distances to estimate total counts and found that blue clusters extend further out than red clusters, as is typical of nearby GC systems. Unfortunately, we cannot extrapolate the missing cluster metallicities. Instead, we can reduce our model samples to match the observational setup as close as possible. Since we have no information on the spatial distribution of model clusters within individual halos, all we can do is remove clusters brought in by satellite halos and presumably deposited outside $8\kpc$ from the center. Most of them were already excluded by construction: when extracting merger trees from the MM-II database, we did not include any satellite halos within the virial radius of the central halo at $z=0$.

In addition, there could be satellites in the central halo merger tree at high redshift that have not had sufficient time to migrate toward the center and deposit their GCs within the ACS field. To identify such satellites, we estimate the dynamical friction timescale of all halos in the tree based on their mass and position information in the MM-II database. The expression for the inspiral time is derived from the \citet{chandrasekhar43} formula, with numerical corrections based on cosmological simulations, e.g., Equation~(5) of \citet{boylan-kolchin_etal08}: %Here, we consider two results from Equation~(5) of \citet{boylan-kolchin_etal08} and Equation~(1) of \citet{jiang_etal10}:
\begin{equation}\label{eq:df_boylan}
\frac{\tau_{\rm merge}}{\tau_{\rm dyn}} = 0.216\, \frac{(M_{\rm host}/M_{sat})^{1.3}}{\ln(1+M_{\rm host}/M_{sat})}\, e^{1.9\eta}\, \left(\frac{r}{r_{\rm vir}}\right),
\end{equation}
where $\eta=j/j_c$ is the orbital circularity and $r/r_{\rm vir}$ is the distance between host and satellite halo in the unit of host virial radius. The most likely value of the circularity is $\eta\approx0.5$, based on the orbital analysis in the simulations of \citet{boylan-kolchin_etal08}. We discard the GCs of each satellite halo at every epoch since $z=2$ (at $z>2$ all satellites have sufficiently short merging time) that had inspiral time longer than the available time until $z=0$. An alternative expression for $\tau_{\rm merge}$ is given by \citet{jiang_etal10}, but the result of this satellite removal process is very similar in both cases.

This procedure affects only 5\%-10\% of GCs, mainly in the metallicity range of $-2.3 < \feh < -1.0$, and does not remove clusters with higher metallicity. Such small changes can be understood intuitively. First, massive satellites have short merger timescales so that they are almost guaranteed to merge. Small satellites with longer inspiral times contain few GCs and cannot affect the overall metallicity distribution. Second, the metallicity of these discarded GCs is roughly in the middle of the distribution, not the poorest which come from the high redshift galaxies and not the richest which come from the central halos on the main branch of the merger tree. This dip in the middle of the metallicity distribution helps to sharpen the appearance of bimodality, although the effect is small. After the exclusion of the satellite GCs, the best fit parameters of Model 1 remain the same as those in Table~\ref{tab:other_models}, with a similar goodness value $G_{0.01}=0.38$.

In the remainder of this paper we present the results for our full model samples, because the exclusion correction is small and involves additional steps (such as the dynamical friction time estimate) that unnecessarily complicate the model.

\subsection{Metallicity Distribution}

Figure~\ref{fig:four-samples} shows the calculated GC metallicities within four representative MM-II halos for our fiducial Model 1, and the observed samples of the Virgo galaxies matched to these halos. This figure illustrates that the model produces realistic GC populations with the multi-modal metallicity distribution. Quantitatively, both the height and location of the blue and red peaks match well with the corresponding observations, in the full halo mass range from $\Mh \sim 10^{12}$ to $10^{14}\Msun$. 

Using the Gaussian Mixture Modeling code \citep{muratov_gnedin10}, we determine the metallicities of the two modal peaks for all 20 halos, as well as for the Virgo galaxies. We also fit the relation between the stellar mass and the peak metallicity of the blue and red populations, $\rm Z \propto M_*^{\gamma}$. Figure~\ref{fig:peaks} shows that the model matches the observed locations of both peaks and follows the weak trend of increasing peak metallicity with galaxy stellar mass. The best-fit slopes are $\gamma=0.24\pm0.17$ and $\gamma=0.01\pm0.21$ for the metal-poor and metal-rich populations, respectively.

\subsection{Mass Distribution}

\begin{figure}[t]
  \vspace{-0.3cm}
\hspace*{-0.4cm}\includegraphics[width=1.15\hsize]{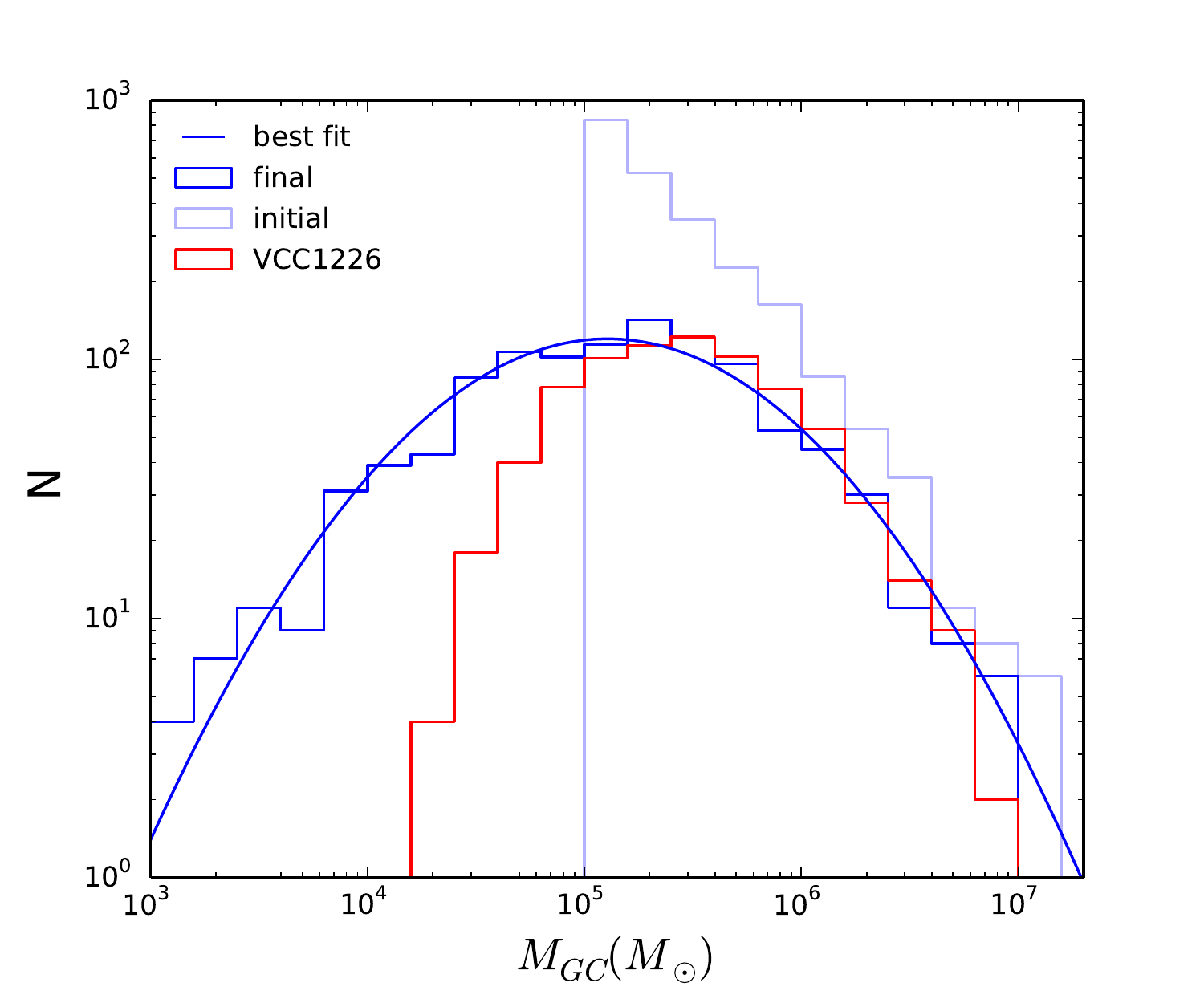}
  \vspace{-0.3cm}
\caption{\small Dynamical evolution of the GC mass function from an initial power law (light blue histogram) to the current peaked distribution (blue histogram, with the overplotted log-normal fit), in the fiducial Model~1 for a halo of $6.9\times 10^{13}\Msun$.  The mass function of the GC system in VCC~1226 is shown for comparison (red histogram).  A sharp drop-off at low mass is likely due to incompleteness of the observed sample.  The KS test comparison of the model and data shows that they are consistent; $p_{KS} \approx 2\%$ for clusters with $M > 10^{5}\Msun$.}
  \vspace{0.1cm}
    \label{fig:mass-dis}
\end{figure}

In this section we examine the mass distribution of model GCs. Figure~ \ref{fig:mass-dis} shows the initial and final mass functions of the largest halo with $\Mh = 6.9\times 10^{13}\Msun$. The dynamical erosion of the GC system turns the initial power-law shape to a peaked shape, which is consistent with the theoretical expectations and observations. We fit the final mass distribution by a conventional log-normal function:
\begin{equation}
  \frac{dN}{d\log M}=\frac{1}{\sqrt{2\pi}\sigma_M}\exp\left[-\frac{(\log M-  \overline{\log M})^2}{2\sigma_M^2}\right],
\end{equation}
with best-fit parameters $\overline{\log M}=5.10$ and $\sigma_M=0.69$. The GC mass function within VCC 1226, obtained from the $(g-z)$ color and the color-dependent mass-to-light ratio, is overplotted in the same figure. The observed distribution is similar to the modeled one for GCs more massive than $10^{5}\Msun$. For lower mass clusters, the observed distribution falls off sharply. It is most likely due to incompleteness of the flux-limited sample. Deeper observations are needed to investigate whether the true mass function in VCC 1226 is described by a similar log-normal.

\subsection{Best-fit Parameter Sets for Other Models}

In Section~\ref{sec:other_models}, we introduced alternative models by modifying particular assumptions of Model 1. We have repeated the model comparison procedure to explore the parameter space for each of these models. Their best-fit parameters are listed in Table~\ref{tab:other_models}. 

\begin{figure}[t]
  \vspace{0cm}
\hspace*{-0.35cm}\includegraphics[width=1.05\hsize]{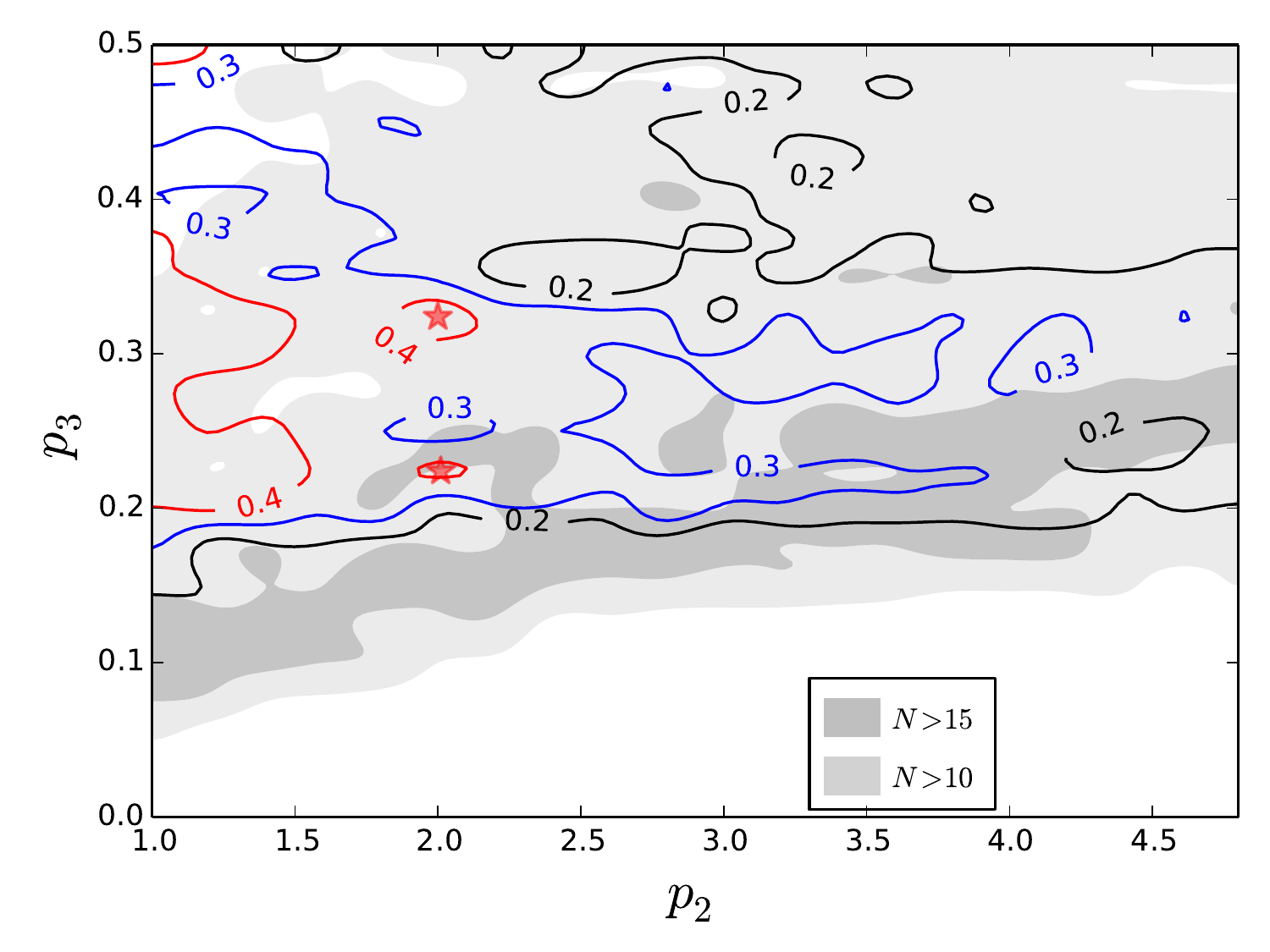}
\hspace*{-0.35cm}\includegraphics[width=1.05\hsize]{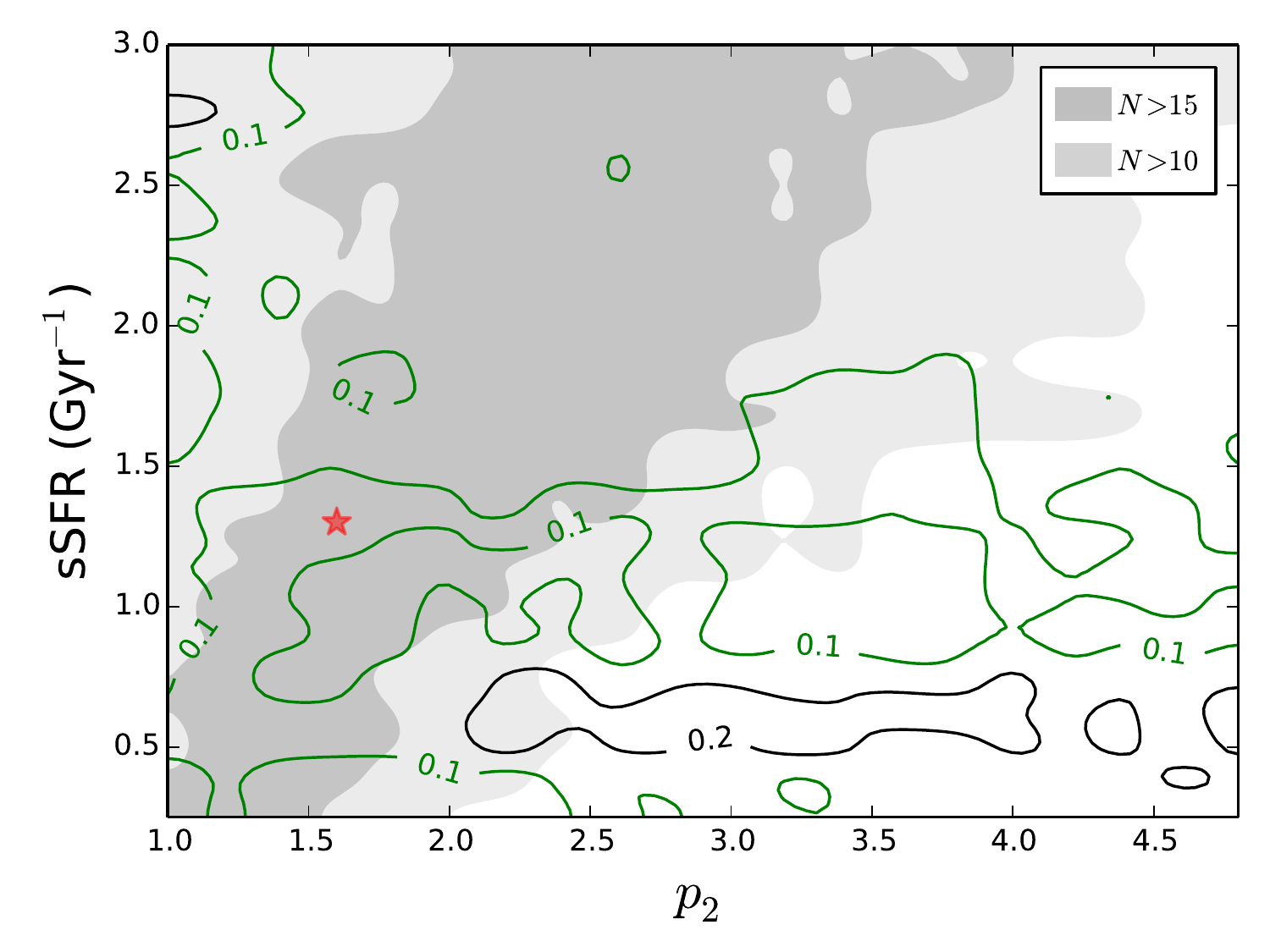}
  \vspace{-0.3cm}
\caption{\small Same as Figure~\ref{fig:p2p3_model1}, but for the alternative Model 3 (top panel) and Model 4 (bottom panel). Best-fit parameters for these models are given in Table~\ref{tab:other_models}.}
  \vspace{0cm}
  \label{fig:p2p3_alter}
\end{figure}

In Model 2, the no-evolution of the MMR causes a systematic shift of model GCs to higher metallicity, which cannot match the observed values of the red and blue modal peaks. As we show in Section~\ref{sec:age_metal}, the metal-poor population, in general, is formed at high redshift around $z \approx 4-8$, when the MMR evolution would lower $\feh$ by about 0.5~dex (see Equation~(\ref{eq:metal_evo})). On the other hand, the most metal-poor GCs formed in the smallest halos with stellar mass $M_* \sim 2\times 10^{6}\Msun$ can only reach $\feh \approx -1.7$ without MMR evolution. Thus, the observed GCs with $\feh = -2.5$ cannot be recovered even after adding the scatter $\sigmamet=0.2$. The "goodness" statistic for Model 2 is quite low (2\%) even when we vary all the parameters in a wide range. Such a significant difference from the results of Model 1 may imply that the moderate evolution of the MMR with cosmic time is favored by the GC systems of massive early-type galaxies. However, at the current stage it is difficult to constrain the exact amount of the evolution because of other intrinsic uncertainties in the model. 
%\textbf{Qualitatively, the moderate MMR evolution we applied here together with its scatter $\sigmamet$ makes it possible to reproduce both subpopulations of GCs.}

Is the metallicity scatter necessary?  An alternative way of populating the metal-poor tail without the scatter is to apply stronger MMR evolution. To investigate this possibility, we set $\sigma_{\rm met}=0$ and vary the coefficient in the second term of Equation~(\ref{eq:metal_evo}) from 0.216 to 0.5. However, for all of these values, the goodness of fit is low ($G_{0.01} < 5\%$), which indicates that boosting the MMR evolution alone cannot substitute for the effect of scatter. Stronger MMR evolution creates several problems. First, both blue and red GC populations are shifted to lower metallicity, which leads to incorrect peak positions. At the same time, the metal-rich clusters with $\feh>0$ are difficult to form without the scatter. Second, the scatter not only helps to fill both tails of the distribution ($\feh < - 2$ and $\feh > 0$) but it also regulates the width of the two populations. Without the scatter, the metallicity distribution is more like a sum of delta functions rather than a Gaussian shape, especially for the metal-rich GCs formed by late discrete mergers.

For both Model 3 and Model 4, the contours of goodness $G_{0.01}$ are shown in Figure~\ref{fig:p2p3_alter}. Model 3 has two peaks with $G_{0.01} > 40\%$ (marked by red stars), which are as good as Model 1. However, the peak at $(p_2, p_3 = 2.0, 0.32)$ does not have as high a number of sufficient GC systems. We have determined the red and blue peak metallicities for this model and added them to Figure~\ref{fig:peaks} for comparison with Model 1. The average peak metallicities are similar, but the dispersion in Model 3b is much larger. The combination of these comparisons makes us prefer Model 1 as the fiducial model.

The highest goodness of Model 4 is only $G_{0.01} = 18\%$, which means that fewer than one in five of the galaxy-halo pairs have acceptable metallicity distributions, significantly lower than in both Model 1 and Model 3. The best-fit critical sSFR to trigger cluster formation is $\sSFR_{\rm crit} \approx 1.3\, \Gyr^{-1}$.  We also tried using the critical SFR as the trigger parameter and found it to be even more difficult to reproduce the observations. The low goodness of Model 4 indicates that the major merger scenario may indeed be a dominate formation channel of GCs, at least in our model.

\begin{figure}[t]
\hspace*{-0.4cm}\includegraphics[width=1.08\hsize]{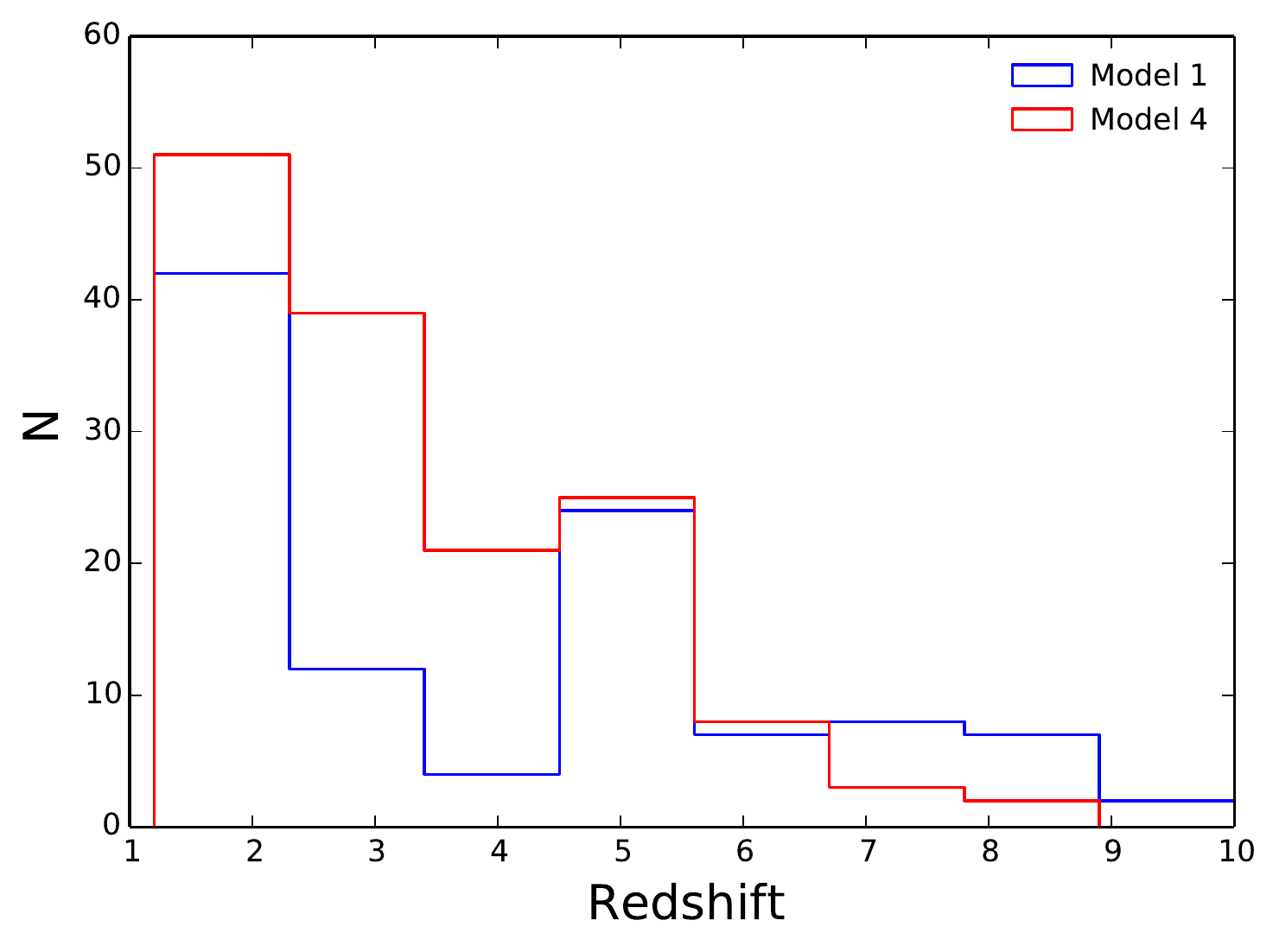}
  \vspace{-0.3cm}
\caption{\small Distribution of GC formation redshift in Model 1 and Model 4 within a $2\times 10^{12}\Msun$ halo.}
  \vspace{0.2cm}
  \label{fig:z_dis}
\end{figure}

\begin{figure}[t]
  \vspace{-0.4cm}
\hspace*{-0.4cm}\includegraphics[width=1.14\hsize]{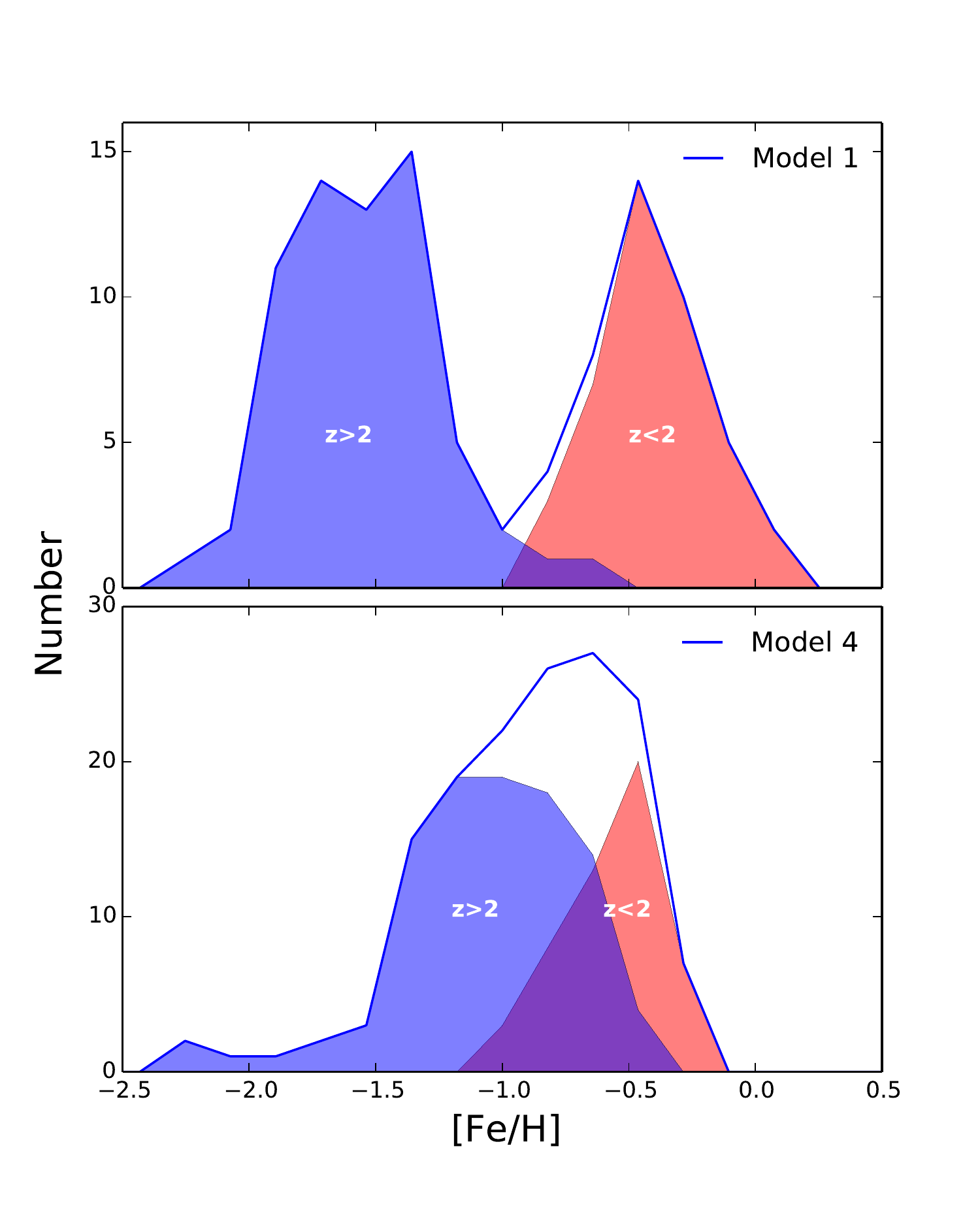}
  \vspace{-0.8cm}
\caption{\small Metallicity distribution of GCs within a $2\times 10^{12}\Msun$ halo for Model 1 ({\it upper panel}) and Model 4 ({\it lower panel}). The samples are split into two groups based on formation redshift: $z<2$ ({\it red shaded}) and $z>2$ ({\it blue shaded}).}
  \vspace{0.2cm}
  \label{fig:m1_m4}
\end{figure}

It should be mentioned that since our GC formation model is based on the merger tree extracted from the MM-II simulation with only 67 outputs along the whole cosmic time, the SFR we derive here is the average between two adjacent outputs. This averaging smoothes out short starburst events. Until we have simulations with high enough time resolution, the short-duration effects cannot be incorporated correctly.
It is still interesting to investigate the differences between results of the merger (Model 1) and starburst (Model 4) scenarios. Figure~\ref{fig:z_dis} shows the formation redshifts of GCs in the two models for a $2\times 10^{12}\Msun$ halo. In Model 4, GC formation activity increases continuously toward relatively low redshift, $z \approx 1-2$. In contrast, Model 1 shows two formation epochs, one at low redshift when the last major merger happened between massive halos and another at higher redshift ($z \approx 5$) when mergers among small halos happened frequently. These differences in the formation history, together with the halo mass growth and MMR evolution, translate into the final GC metallicity distribution. Figure~\ref{fig:m1_m4} illustrates how the metallicity bimodality is produced by discreteness of the late mergers.

In contrast, continuous field star formation, during and between mergers, does not lead to a bimodal metallicity distribution. In order to show it within our framework, we modeled the field metallicity distribution as a mass-weighted sum of stellar populations formed at each simulation output.  We calculated the mass increments of all halos (central and satellite) in the merger tree between successive outputs, converted them to stellar mass using the stellar mass-halo mass relation \citep{behroozi_etal13b}, and evaluated the metallicity of such a stellar population using the evolving MMR (Eq.~(\ref{eq:metal_evo})).  The sum of these contributions roughly represents the metallicity distribution of the field stars. This distribution is clearly unimodal, in agreement with well-known observations \citep[e.g.][]{harris_harris02}. For example, the peak metallicity of a $2\times10^{12}\Msun$ halo is at $\feh \approx -0.22$ and the median is around $\feh \approx -0.28$. In contrast, as we can see in Figure~\ref{fig:peaks}, the two peaks of the GC metallicity distribution are at $\feh \approx -1.54 ~ \& -0.4$ for the metal-poor and metal-rich populations, respectively, with the median at $\feh \approx -1.30$. This comparison shows that, although GCs and field stars are forming concurrently, the gas-rich merger-driven cluster formation filters a bimodal metallicity distribution from an extended unimodal one.

\begin{figure}[t]
  \vspace{0cm}
\hspace*{-0.35cm}\includegraphics[width=1.05\hsize]{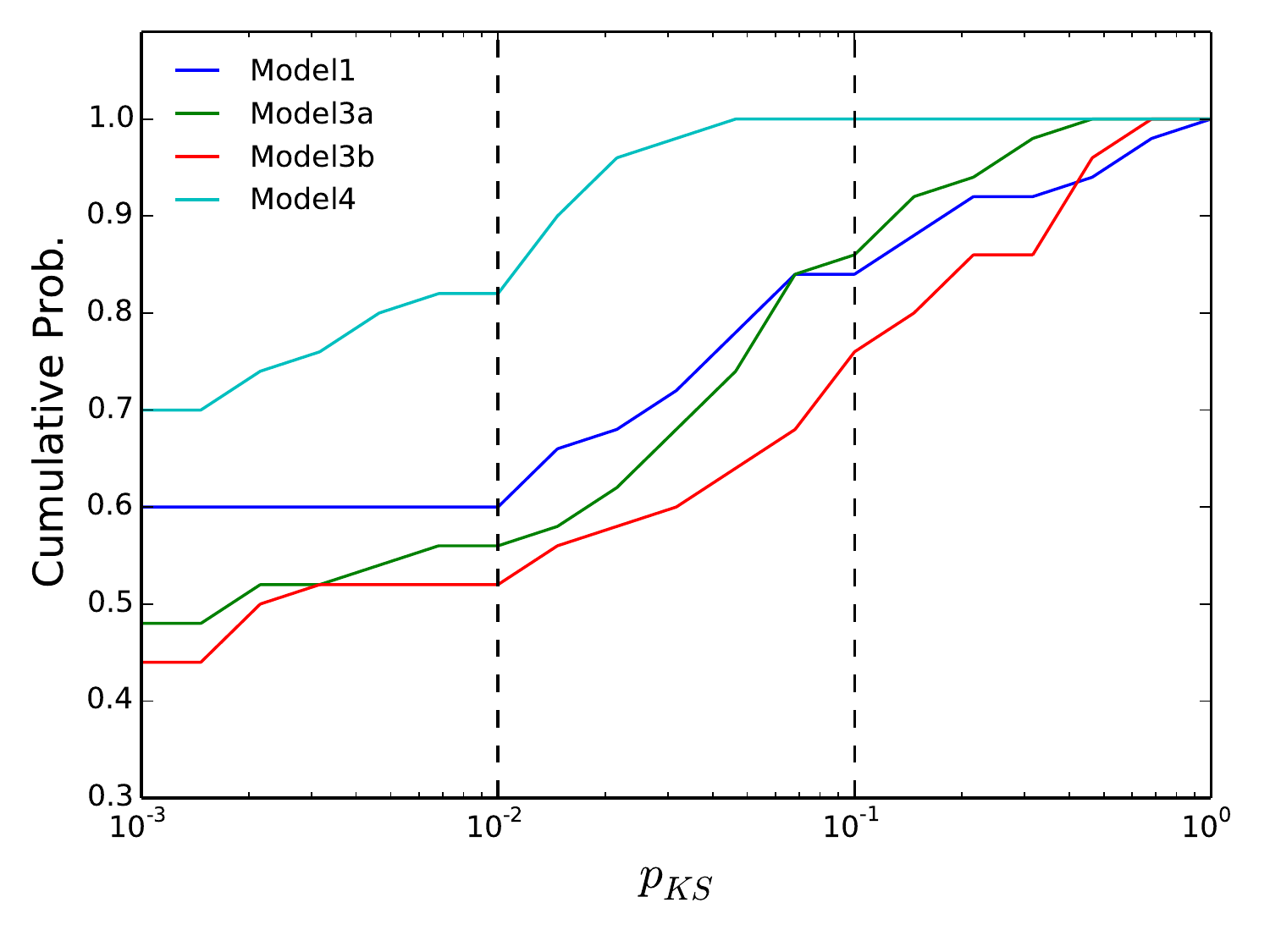}
  \vspace{-0.5cm}
\caption{\small Cumulative distribution of $p$-values of the KS test for the metallicity distribution, for all best-fitting models. The vertical scale is related to the "goodness" parameter as $1-G_{p_{KS}}$. Lower lines have higher "goodness" of fit.}
  \vspace{0.1cm}
  \label{fig:pvalue_models}
\end{figure}

To further investigate the quality of fit of different models, in Figure~\ref{fig:pvalue_models} we present the full cumulative distribution of the KS test $p$-values for the best parameters of each model (Table~\ref{tab:other_models}). Most of the $p$-values of Model 4 are below $10^{-3}$, so that the cumulative probability at $p_{KS} \gtrsim 1\%$ is already far above that of Model 1 and Model 3. The performance of Model 1 and Model 3a is fairly similar, which indicates that our major merger scenario is not too sensitive to the details of the cold gas modeling. Model 3b appears a little better than the other two, but as we discussed above, it cannot reproduce the number of GCs as well and has a larger scatter of the modal peak metallicities (Figure~\ref{fig:peaks}). 

\begin{table}
\begin{center}
\caption{\sc Comparison of Best-fit Model Parameters}
  \label{tab:other_models}
\begin{tabular}{lllll}
\tableline\tableline\\
\multicolumn{1}{l}{Model} &
\multicolumn{1}{l}{$p_2$} &
\multicolumn{1}{l}{$p_3$} &
\multicolumn{1}{l}{$\mathrm{sSFR}$} &
\multicolumn{1}{l}{Goodness}\\
\multicolumn{1}{l}{} &
\multicolumn{1}{l}{} &
\multicolumn{1}{l}{} &
\multicolumn{1}{l}{$(\Gyr^{-1})$} &
\multicolumn{1}{l}{$G_{0.01}$}
\\[2mm] \tableline\\
Model 1  & 2.6 & 0.33 & --  & 0.40 \\
Model 2  & 2.6 & 0.33 & --  & 0.02 \\
Model 3a & 2.0 & 0.22 & --  & 0.46 \\
Model 3b & 2.0 & 0.32 & --  & 0.48 \\
Model 4  & 1.6 & --   & 1.3 & 0.18
\\[2mm] \tableline
\end{tabular}
\end{center}
\vspace{0.1cm}
\end{table}

\begin{figure}[t]
  \vspace{-0.3cm}
\hspace*{-0.4cm}\includegraphics[width=1.12\hsize]{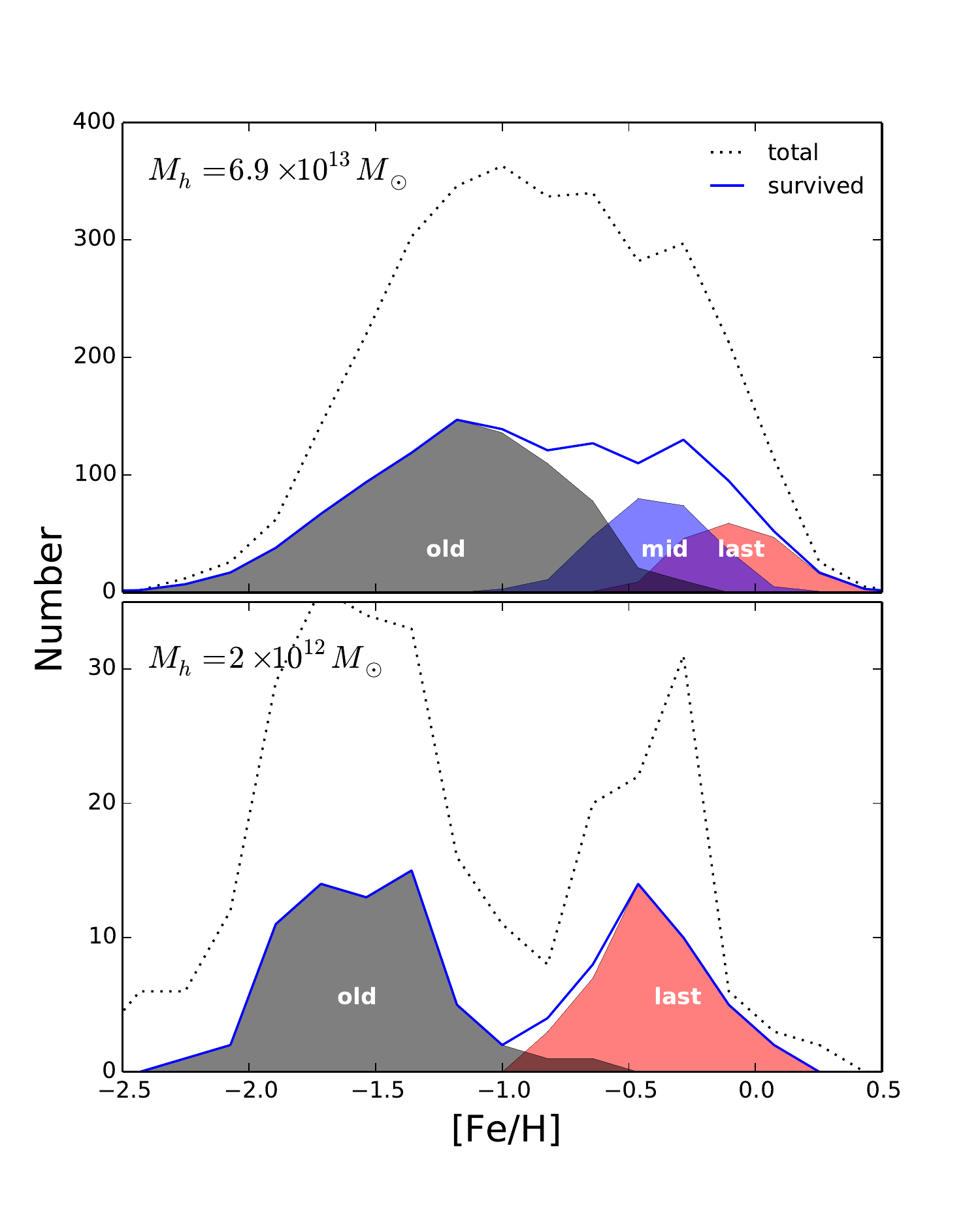}
  \vspace{-0.8cm}
\caption{\small Metallicity distributions of the total (dotted lines) and survived (blue solid lines) GC systems within $6.9\times10^{13}M_\odot$ halo (upper panel) and $2\times10^{12}M_\odot$ halo (lower panel). The distributions are also split by the merger events that produced the clusters: late mergers (red shaded), intermediate mergers (blue shaded), and early mergers (gray shaded).}
  \vspace{0.2cm}
  \label{fig:metal_dist_split}
\end{figure}

\section{Discussion}
  \label{sec:discussion}

\citet{ashman_zepf92} proposed the idea that GCs can be formed in mergers between gas-rich galaxies, since such mergers can perturb the gravitational potential, shock and compress the ISM within the two galaxies, and trigger large-scale starbursts. HST observations have already demonstrated many interacting galaxies with young massive star clusters, whose formation was likely triggered by merging \citep[e.g.,][]{holtzman_etal92, whitmore04, larsen09}. Our best-fitting Model 1 suggests a minimum merger ratio of 1:3 for triggering cluster formation, consistent with this major merger scenario.

\citet{tonini13} proposed a model for metallicity bimodality based on the observed number of clusters as a function of galaxy mass.  She adopted the merger scenario and used Monte Carlo sampling to build merger trees for the progenitor galaxies. She suggested that the origin of GC bimodality is related mainly to the galactic MMR and hierarchical mass assembly history. Using our model, we reach a similar conclusion that the merger history plays a key role. However, to separate the red and blue peaks, the \citet{tonini13} model requires a very strong evolution of MMR, such that $\feh$ increases by 0.5~dex at high mass ($M_* \sim 10^{11.5}\Msun$) and up to 1.5~dex at low mass ($M_* \sim 10^{9}\Msun$), between $z=3.5$ and $z=0$.  The available observations discussed in Section~\ref{sec:metal} support much smaller changes of $\feh$ at a given stellar mass.

Our model also relies on the evolution of MMR, but the evolution we need is more moderate ($\approx 0.3$~dex). The key to separating the metal-poor and metal-rich subpopulations in our model is mainly due to the differentiation of cluster hosts. The metal-poor GCs come preferentially from the early mergers among small halos with lower metallicity, while the metal-rich GCs come from the late mergers between massive halos, which in turn have higher metallicity.

To demonstrate this effect, we select two halos with the highest and lowest mass ($6.9\times 10^{13}\Msun$ and $2\times 10^{12}\Msun$, respectively) and separate their GC systems by the merger epoch in which they were produced.  Figure~\ref{fig:metal_dist_split} shows the result for the best-fit parameters of Model 1.  Although the dynamical destruction significantly reduces the number of surviving clusters, the shape of the metallicity distribution does not change much from that imprinted at birth.  On the other hand, the merger events that produced the clusters leave a clear mark. The metal-rich GCs are mainly produced by the most recent merger between massive halos, while the collection of early mergers among less massive halos contributes the bulk of the metal-poor clusters.

\begin{figure}[t] 
  \vspace{0cm}
\hspace*{-0.35cm}\includegraphics[width=1.1\hsize]{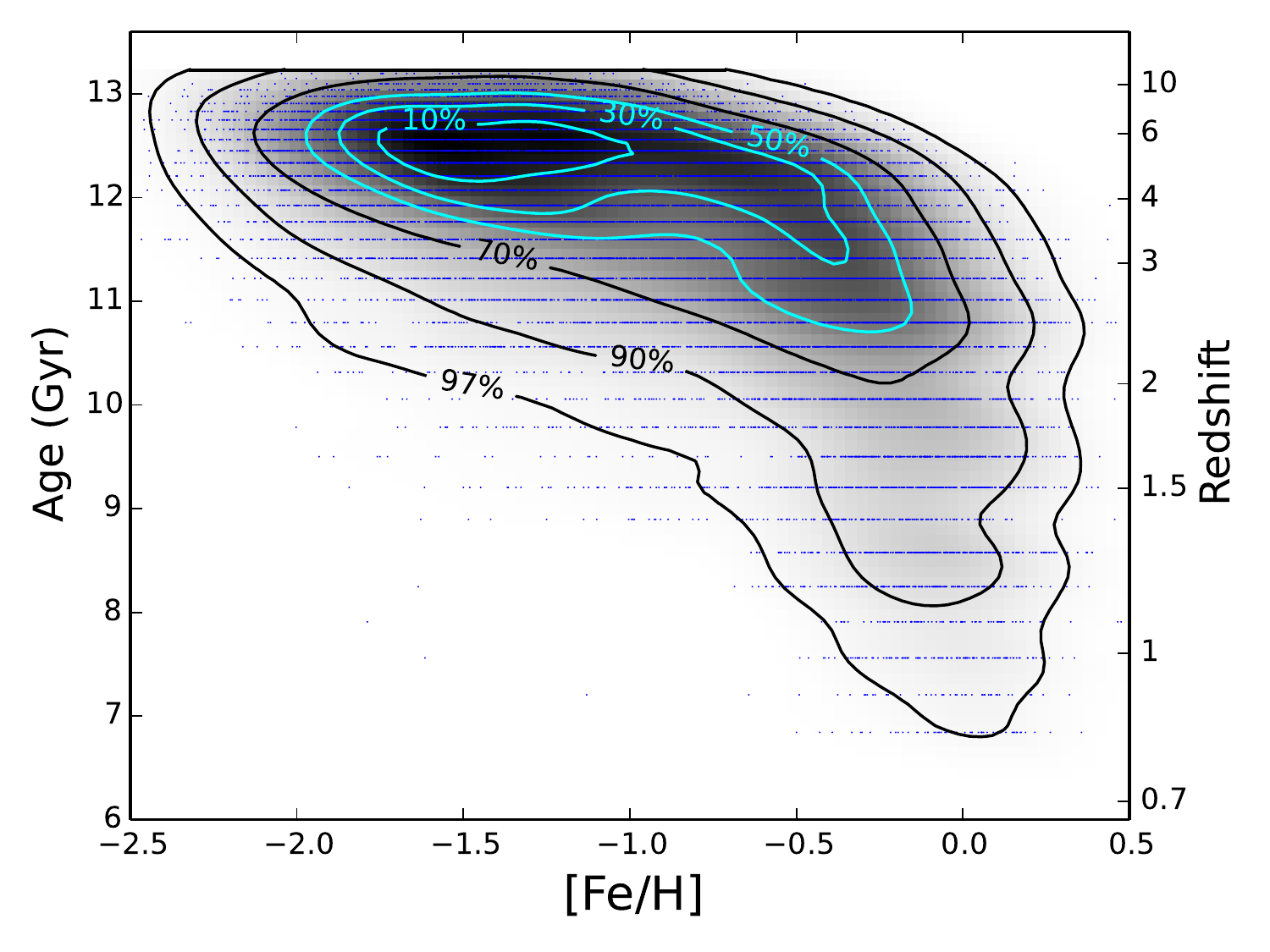}
  \vspace{-0.3cm}
\caption{\small Age-metallicity distribution of model clusters from all 20 halos, with the best-fit parameters of Model 1. Each blue dot represents one model GC. The labeled percentage levels represent the fraction of GCs enclosed within the corresponding contour. Redshift corresponds to the cluster formation epoch.}
  \vspace{0.1cm}
   \label{fig:age_metal}
\end{figure}

\begin{figure}[t] 
  \vspace{0cm}
\hspace*{-0.35cm}\includegraphics[width=1.1\hsize]{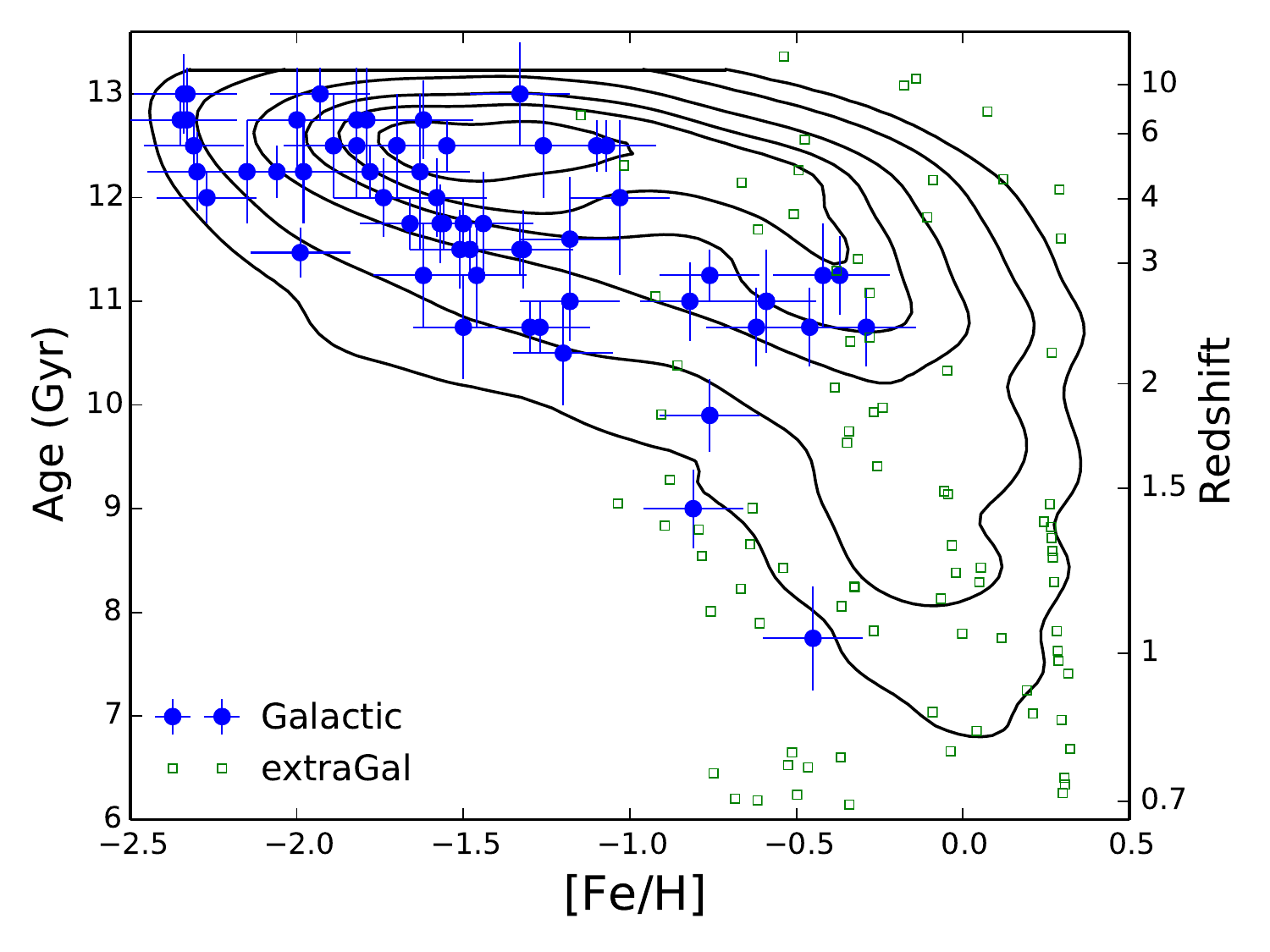}
  \vspace{-0.3cm}
\caption{\small Age-metallicity relation of Galactic GCs (filled circles with error-bars; from \citealt{leaman_etal13} and other sources, see the text) and extragalactic GCs (open squares; from \citealt{georgiev_etal12}).  Overlaid contours are the same as in Figure~\ref{fig:age_metal}, for our model clusters.}
  \vspace{0.1cm}
   \label{fig:age_metal_obs}
\end{figure}

\subsection{Age-metallicity Relation}\label{sec:age_metal}

Absolute ages of GCs can be determined using isochrone fitting of the H-R diagram, which requires resolved observations of individual stars. Until recently, the age measurements of Galactic GCs did not show a significant correlation between age and metallicity \citep[e.g. ][]{forbes_bridges10}. New deep HST/ACS data reveal some intriguing trends of decreasing age with increasing metallicity \citep{dotter_etal11, vandenberg_etal13, leaman_etal13}.  In our model we have the full formation history of all GCs, which allows us to investigate any possible age-metallicity trends.  Figure~\ref{fig:age_metal} shows a stack of all model GCs within the 20 halos in the fiducial Model 1. Although the majority of clusters are old, there is a significant tail of metal-rich clusters that are younger by up to $5\Gyr$. The bulk of metal-poor clusters are formed as early as redshift $z=4-6$, but the metal-rich clusters are formed over an extended epoch continuing to $z \approx 1$. This shape of the age-metallicity distribution is one of the robust predictions of our model.

Hints of the age-metallicity relation were already present in the MG10 model (see their Figure~8). Here we quantify it with larger samples of clusters, multiple independent realizations of the mass assembly history, and better galactic scaling relations. Figure~\ref{fig:age_metal_obs} compares the model relation with the existing age measurements of the Galactic GCs, collected by \citet{leaman_etal13}. We also add clusters from three nearby early-type galaxies with photometrically derived ages \citet{georgiev_etal12}. Despite the large scatter and individual observational errors, the model trend is supported by these data remarkably well.

A thorough interpretation of this plot requires further study.  It is likely that the turnover of the age-metallicity relation from the old metal-poor clusters to the younger metal-rich clusters occurs at different metallicity in galaxies of different mass. For example, in the MG10 model tuned for the Galactic GCs, the turnover is around $\feh \approx -0.8$, whereas in our current model tuned for massive elliptical galaxies it is around $\feh \approx -0.4$. In addition, the ages of the extragalactic clusters are determined with a different method and different fidelity than those of the Galactic GCs.  Nevertheless, the emerging age-metallicity relation of GCs is tantalizing and invites further accurate measurements of cluster ages in extragalactic systems.

\section{Summary and Conclusions}
  \label{sec:summary}

We have constructed a model of cluster formation, incorporating the halo merger trees from the MM-II cosmological N-body simulation, to investigate the origin of GC systems in massive early-type galaxies. We include the empirical galactic scaling relations, such as the stellar mass-halo mass relation, stellar mass-gas mass relation, and stellar MMR. These come either from direct observations or from the empirical abundance matching technique. We test the scenario in which clusters are formed as a result of major mergers of gas-rich galaxies. By matching the masses of our selected halos with the galaxies in the Virgo cluster, we compare the metallicity distributions of modeled and observed GCs and thus constrain the model parameters. We have also tested alternative models in order to examine the sensitivity of our results to various adopted prescriptions. Our main conclusions are listed below:

\begin{itemize}
\item Our fiducial model can successfully reproduce both the number and the metallicity distribution of GCs within a large range of halo masses from $2\times 10^{12}\Msun$ to $6.9\times 10^{13}\Msun$. The metallicity distribution appears to have a bimodal shape, and the metallicities of the blue and red peaks are consistent with those observed in the Virgo galaxies. 

\item The fiducial model requires a minimum merger ratio of 1:3 to trigger cluster formation. This ratio is consistent with the theoretical expectation of a major merger.

\item A detailed analysis of the formation history of GCs reveals that the bimodality arises from different merger epochs and host galaxy masses: the metal-rich population is produced by late mergers between massive halos, while the metal-poor population is produced by early mergers among less massive halos.

\item The model predicts a robust age-metallicity relation of GCs, which can be falsified by further observations. While the bulk of metal-poor clusters are very old, the metal-rich clusters are progressively younger, by up to $5\Gyr$.

\item When the evolution of the galaxy MMR with cosmic time is turned off, the model GC metallicity distribution shifts to higher $\feh$ and the bimodal distribution disappears. This suggests that the evolution of MMR is necessary in our model.

\item The evolution of the cold gas fraction within galaxies at high redshift is largely unconstrained by current observations. We use different methods to parameterize this evolution and find that the best-fitting model results for the GC number and metallicity distributions are insensitive to the details of the adopted prescriptions, within the considered range.

\item We also challenged our major merger scenario and tested an alternative starburst scenario, which required a minimum sSFR to trigger cluster formation. Because of the smooth behavior of the average sSFR derived from the abundance matching, the alternative model fails to reproduce the observed metallicity distribution as well as the merger model.
\end{itemize}

\acknowledgements
We thank Peter Behroozi, Paul Goudfrooij, Evan Kirby, Andrey Kravtsov, Diederik Kruijssen, and Alexander Muratov for helpful discussions. We are especially grateful to Ryan Leaman, Eric Peng, and Joachim Vanderbeke for sending us their data, and to Jeremy Bradford and Marla Geha for sharing their results before publication. This work was supported in part by the National Science Foundation under grant PHYS-1066293 and the hospitality of the Aspen Center for Physics, Kavli Institute for Theoretical Physics in Santa Barbara, and Kavli Institute for Cosmological Physics in Chicago.  O.Y.G. is supported in part by NASA through grant NNX12AG44G.

\makeatletter\@chicagotrue\makeatother

\bibliographystyle{apj}
\bibliography{gc,hui_gc}

\end{document}